\documentclass[apj]{emulateapj}
\slugcomment{Accepted to the Astrophysical Journal on March 31, 2017}

\usepackage[normalem]{ulem}

\usepackage{graphicx}
\usepackage{natbib}

\shorttitle{Molecular gas dominated 50~kpc ram pressure stripped tail}
\shortauthors{J\'achym et al.}

\begin{document}
\title{
Molecular gas dominated 50~kpc ram pressure stripped tail\\ of the Coma galaxy D100\altaffilmark{*}}
\author{Pavel J\'achym\altaffilmark{1}, Ming Sun\altaffilmark{2}, 
Jeffrey D. P. Kenney\altaffilmark{3}, Luca Cortese\altaffilmark{4}, 
Fran\c coise Combes\altaffilmark{5}, Masafumi Yagi\altaffilmark{6,7}, 
Michitoshi Yoshida\altaffilmark{8}, Jan Palou\v s\altaffilmark{1}, and 
Elke Roediger\altaffilmark{9}}
\altaffiltext{*}{Based on observations made with the IRAM 30m 
Telescope. IRAM is supported by INSU/CNRS (France), MPG (Germany), and 
IGN (Spain).}
\altaffiltext{1}{Astronomical Institute, Czech Academy of Sciences, Bo\v cn\'i II 1401, 14100, Prague, Czech Republic, \email{jachym@ig.cas.cz}}
\altaffiltext{2}{Department of Physics, University of Alabama in Huntsville, 301 Sparkman Drive, Huntsville, AL 35899, USA}
\altaffiltext{3}{Department of Astronomy, Yale University, 260 Whitney Ave., New Haven, CT 06511, USA}
\altaffiltext{4}{International Centre for Radio Astronomy Research, The University of Western Australia, 35 Stirling Hwy, Crawley, WA 6009, Australia}
\altaffiltext{5}{Observatoire de Paris, LERMA, PSL, CNRS, Sorbonne Univ. UPMC, and College de France, F-75014, Paris, France}
\altaffiltext{6}{Optical and Infrared Astronomy Division, National Astronomical Observatory, Mitaka, Tokyo 181-8588, Japan}
\altaffiltext{7}{Graduate School of Science and Engineering, Hosei University, 3-7-2, Kajinocho, Koganei, Tokyo, 184-8584, Japan}
\altaffiltext{8}{Hiroshima Astrophysical Science Center, Hiroshima University, 1-3-1 Kagamiyama, Higashi-Hiroshima, Hiroshima 739-8526, Japan}
\altaffiltext{9}{Milne Centre for Astrophysics, Department of Physics \& Mathematics, University of Hull, Hull, HU6 7RX, United Kingdom}

\begin{abstract}
We have discovered large amounts of molecular gas, as traced by CO 
emission, in the ram pressure stripped gas tail of the Coma cluster 
galaxy D100 (GMP~2910), out to large distances of about 50~kpc. D100 
has a 60~kpc long, strikingly narrow tail which is bright in X-rays and 
H$\alpha$. Our observations with the IRAM 30m telescope reveal in total 
$\sim 10^9~M_\odot$ of H$_2$ (assuming the standard CO-to-H$_2$ 
conversion) in several regions along the tail, thus indicating that 
molecular gas may dominate its mass. Along the tail we measure a smooth 
gradient in the radial velocity of the CO emission that is offset to 
lower values from the more diffuse H$\alpha$ gas velocities. Such a 
dynamic separation of phases may be due to their differential 
acceleration by ram pressure. D100 is likely being stripped at a high 
orbital velocity $\gtrsim 2200$~km\,s$^{-1}$ by (nearly) peak ram 
pressure. Combined effects of ICM viscosity and magnetic fields may be 
important for the evolution of the stripped ISM. We propose D100 has 
reached a continuous mode of stripping of dense gas remaining in its 
nuclear region. D100 is the second known case of an abundant molecular 
stripped-gas tail, suggesting that conditions in the ICM at the centers 
of galaxy clusters may be favorable for molecularization. From 
comparison with other galaxies, we find there is a good correlation 
between the CO flux and the H$\alpha$ surface brightness in ram 
pressure stripped gas tails, over $\sim 2$~dex.
\end{abstract}


\section{Introduction}
The environments of galactic tails formed by ram pressure stripping 
(RPS) are likely different from typical environments of galaxy disks. 
The interstellar matter (ISM) stripped from galaxies infalling into 
clusters by dynamical pressure of the intra-cluster medium (ICM) that 
fills up the space in between galaxies presumably mixes with the 
surrounding ICM. Due to various competing thermodynamic processes a 
spectrum of temperatures and densities develops in the wakes of 
stripped galaxies. Ram pressure stripping as an efficient 
hydrodynamical mechanism of ISM removal from galaxies in clusters, as 
well as the fate of the stripped ISM, has been studied both 
observationally and theoretically \citep[][and others]{gunn1972, 
nulsen1982, cowie1977, larson1980, giovanelli1983, cayatte1990, 
kenney1999, koopmann2004, chung2009, vollmer2001, roediger2005, 
jachym2007, kapferer2009, tonnesen2011, tonnesen2012}.

In the nearby, $M_{\rm dyn}\sim 10^{14}~M_\odot$, Virgo cluster, many 
short off-disk, mostly \ion{H}{1} features are known \citep{kenney2004, 
chung2009, abramson2011, kenney2014}. While there are also examples of 
longer ram pressure stripped gas tails \citep[length~$\sim (1-2)\times 
D_{25}$;][]{oosterloo2005, chung2007, boselli2016}, most of the gas 
missing in the Virgo galaxies has not been revealed in the 
intra-cluster space \citep[e.g.,][]{vollmer2007}. In the more distant 
cluster A1367, with about 5-times the Virgo mass, several long RPS 
tails were observed in \ion{H}{1} \citep{scott2010, scott2012}.

In more massive clusters ($M_{\rm dyn}\sim 10^{15}~M_\odot$), such as 
Coma or Norma, many examples of long (length~$\gtrsim 3\times D_{25}$), 
clearly RPS tails are observed in (1) diffuse H$\alpha$ 
\citep{gavazzi2001, cortese2006, cortese2007, sun2007, yagi2007, 
yagi2010, yoshida2004, yoshida2008, fossati2012}; (2) X-rays 
\citep{wang2004, finoguenov2004, machacek2005, sun2005, sun2006, 
sun2010}. Some of the tails are observed in multiple wavelengths 
(mostly X-rays + H$\alpha$), such as ESO~137-001, NGC~4848 and GMP2910, 
but also NGC~4388 in Virgo. 

Rather surprisingly, regions of young star formation have been revealed 
in some of the tails, either in H$\alpha$ or UV \citep{kenney1999, cortese2006, 
sun2007, yoshida2008, smith2010, hester2010, yagi2013, ebeling2014}. 
This indicates that while ram pressure generally suppresses star 
formation in the disks, it may induce new star formation in the 
stripped medium. Dense molecular clouds are the principal sites of star 
formation in galaxies. Observed ongoing star formation in some of the 
tails is thus suggestive of the presence of molecular gas. 

Indeed, an abundant cold molecular component was discovered for the 
first time in the tail of the Norma cluster galaxy ESO~137-001 
\citep{jachym2014}. The detected amounts of cold molecular phase ($\sim 
10^9~M_\odot$) are similar to those of hot ionized phase observed in 
the tail, suggesting that molecular gas may form a substantial fraction 
of the multi-phase gas. Moreover, in the tail of ESO~137-001, the 
observed gas components, including the rich molecular phase, together 
nearly account for the missing gas in the disk. Other tails were 
searched for molecular content: In the Virgo cluster, only upper limits 
on CO emission were measured in the tail of IC~3418 \citep{jachym2013}, 
while several regions with molecular gas were recently discovered at 
large distances in the tail of NGC~4388 \citep{verdugo2015}, but 
corresponding only to a small fraction of total mass of the gas tail. 
Also, in the Virgo cluster, regions of off-disk CO emission were 
detected in NGC~4522, \citep{vollmer2008}, as well as in the long 
H$\alpha$ trail that connects M86 with NGC 4438 \citep{dasyra2012}.

Formation and survival of molecular gas in the tails of ram pressure 
stripped galaxies, in the surroundings of the hot ICM, is an 
interesting problem worth of further observational and numerical 
efforts. Central regions of galaxy clusters have revealed that cold 
molecular gas may exist in the long filaments of cooling cores 
\citep[e.g.,][]{salome2011}. The central region of the Coma cluster, 
the most massive and most X-ray luminous cluster at $z<0.025$, is an 
ideal laboratory for studying the hydro-dynamic effects of the 
surrounding ICM on galaxies as well as the fate of the stripped gas. 
The Coma cluster has the richest optical data among nearby massive 
clusters, already with more than 20 late-type galaxies with one-sided 
star-forming or ionized gas tails \citep{smith2010, yagi2010}, several 
of which are also bright in soft X-rays. One of the best galaxies to 
study is D100, a galaxy near the Coma core with a remarkable ram 
pressure gas stripped tail.

\subsection{D100 (GMP~2910)}
\begin{table}
\centering
\caption[]{Parameters of the galaxy D100 (GMP2910, PGC044716, MRK0060 NED01).}
\label{D100}
\begin{tabular}{ll}
\hline
\hline
\noalign{\smallskip}
RA, Dec (J2000)        & $13^{\rm h}00^{\rm m}09^{\rm s}$.14, $+27^{\circ}51'59''.2$\\
type                   & SBab\\
redshift\tablenotemark{a}, $\ V_{\rm helio}$          & 0.01784, 5348~km\,s$^{-1}$\\
$V_{\rm Coma}$         & $-1570$~km\,s$^{-1}$\\
major diameter ($B$-band)  & $24''$\\
major-minor axis ratio & 1.35\\
PA, inclination        & $178.6^\circ$, $\sim 43^\circ$\\
total $B$ ($I)$ mag    & $16.09\pm 0.09$ ($15.23\pm 0.08$)\\
stellar mass\tablenotemark{b} & $2.1\times 10^9~M_\odot$\\
%%rotation velocity      & \\
\noalign{\smallskip}
\hline
\noalign{\smallskip}
\end{tabular}
\tablenotetext{1}{From \citet{yagi2007} and corrected from observer's 
to heliocentric velocity frame.}
\tablenotetext{2}{MEDIAN stellar mass in MPA-JHU SDSS catalog.}
\end{table}

\begin{figure*}[]
\centering
\includegraphics[height=0.32\textwidth]{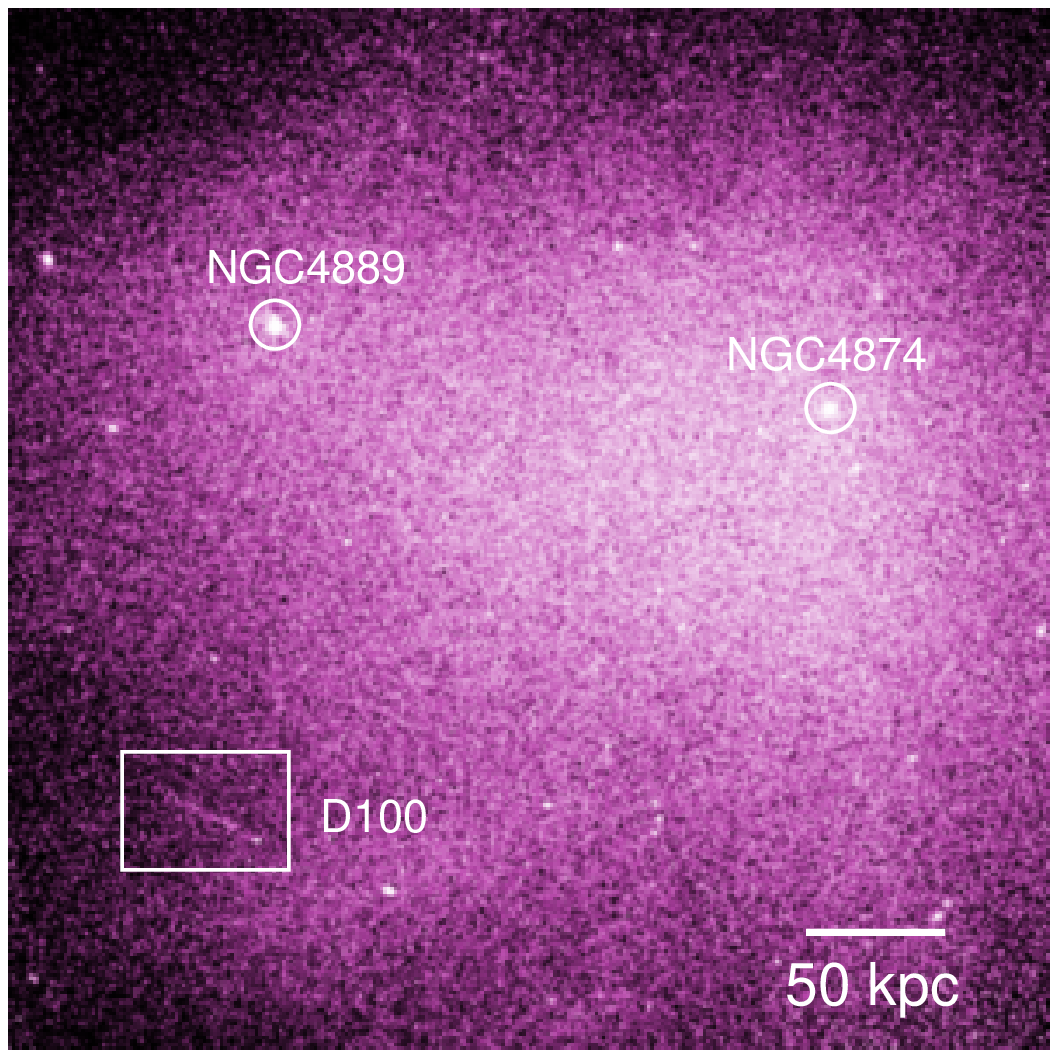}
\includegraphics[height=0.32\textwidth]{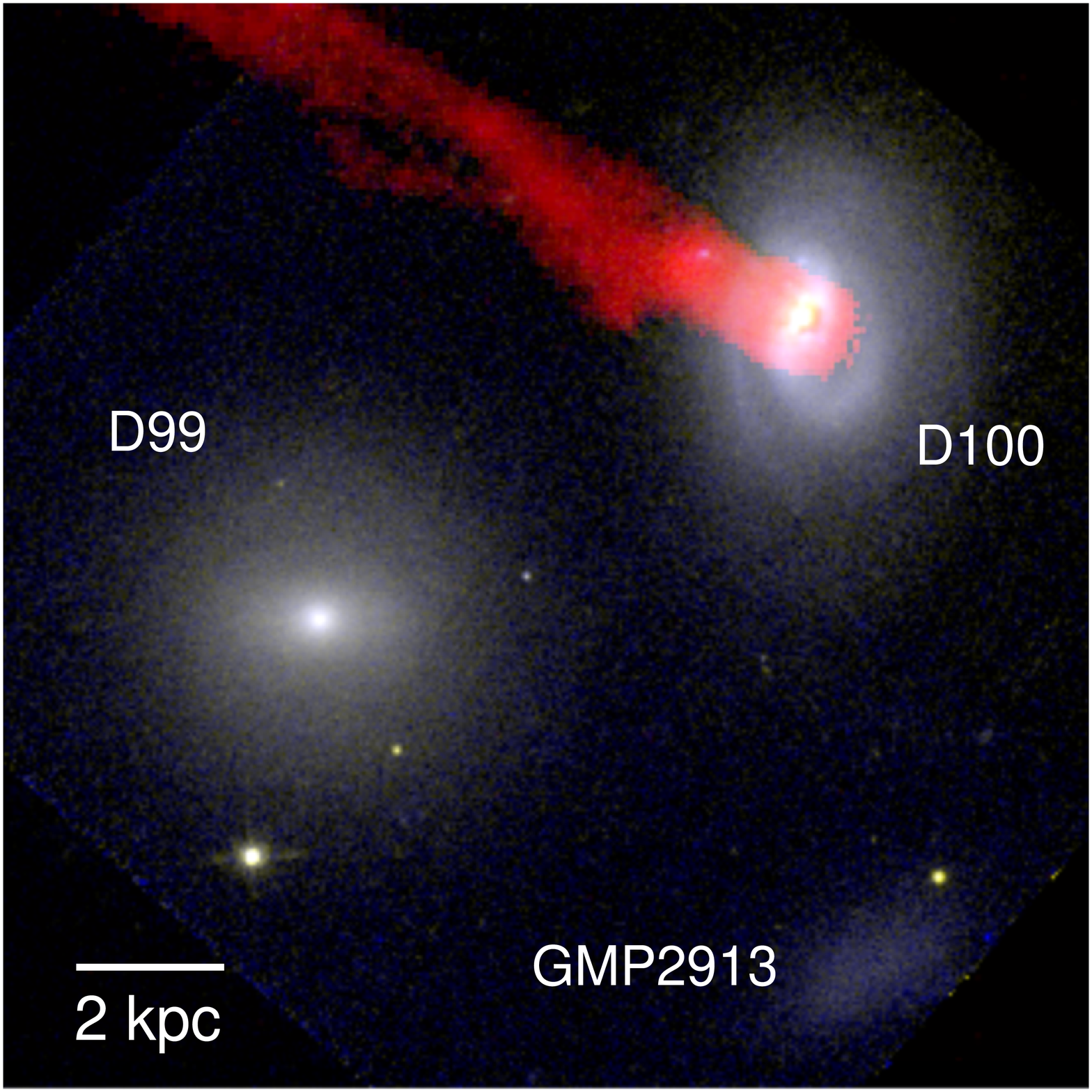}
\includegraphics[height=0.32\textwidth]{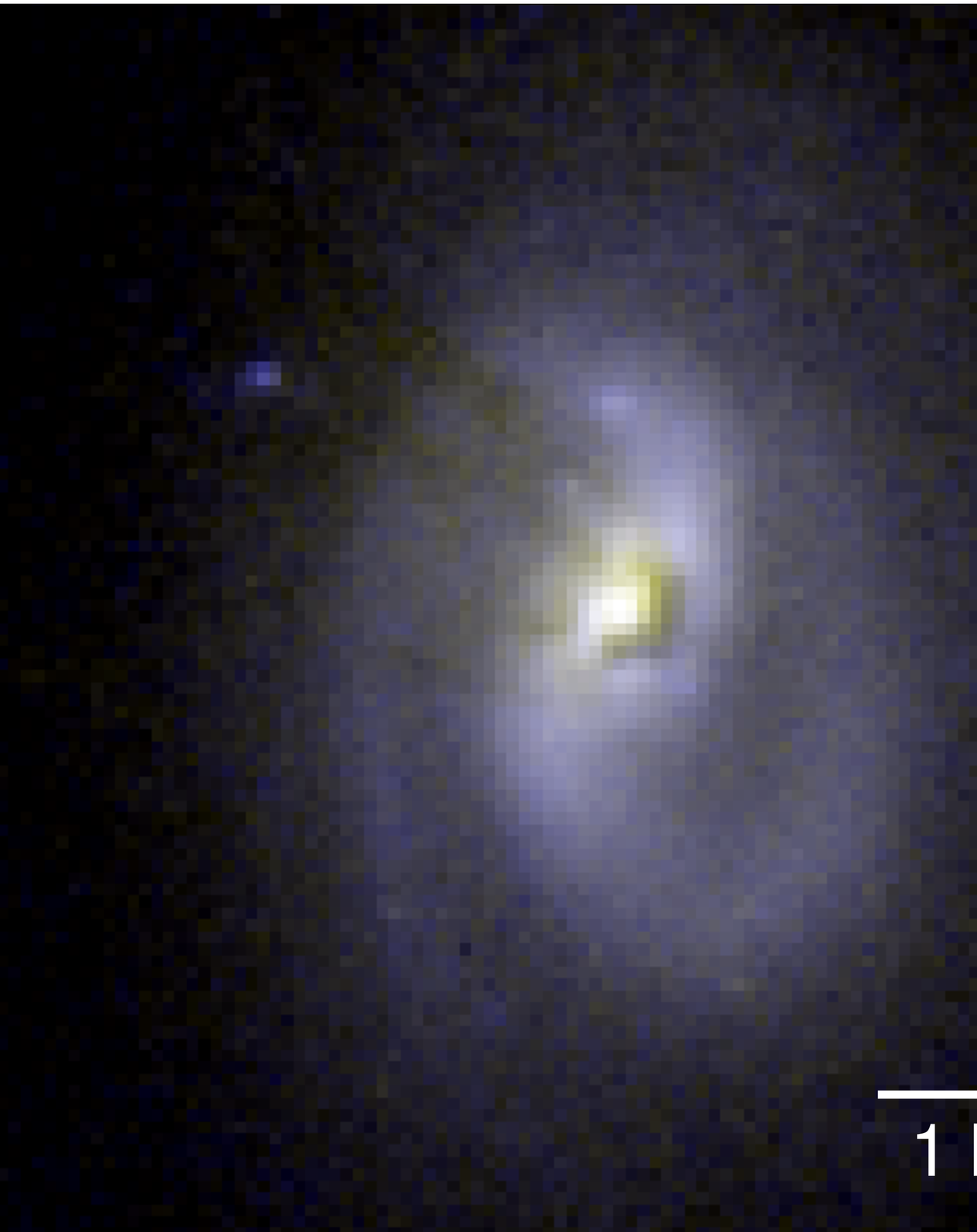}
\caption{
Left: {\it Chandra} view of the central parts of the Coma cluster. 
Positions of two large elliptical galaxies NGC~4889 and NGC~4874 are 
shown with circles, together with the position of D100. The remarkable 
$\sim 48$~kpc long X-ray tail extending to the NE direction from D100 
is clearly visible. Image credit: NASA/CXC/MPE/\citet{sanders2013}. 
Middle: HST view of D100's closest neighborhood: projected to the SE is 
the S0 galaxy D99 (GMP2897) and to the S a weak galaxy GMP2913. The 
Subaru H$\alpha$ tail is overlaid in red \citep{yagi2010}. Right: HST 
WFPC2 zoom on D100 showing prominent dust extinction filaments 
extending from the nucleus, as well as two spiral arms and a central 
bar. The RGB image was created by combining a $B$ (F450W filter, blue), 
an $I$ (F814W filter, red) and a merged $B+I$ (green) images. A F450W 
image was already published by \citet{caldwell1999}. 
}\label{Fig00hst}
\end{figure*}

D100's tail is the straightest, and has the largest length to width 
ratio, of any known ram pressure stripped tail. It is bright in 
multiple wavelengths. Fig.~\ref{Fig003} shows the Subaru deep image of 
the galaxy \citep{yagi2010}. H$\alpha$ emission is coming from long 
($\sim 130''\approx 60$~kpc) and extremely narrow (mostly $\sim 
4.5''\approx 2.1$~kpc) area that connects to the core of the galaxy 
\citep{yagi2007}. The inner parts of the H$\alpha$ tail are also shown 
in the middle panel of Fig.~\ref{Fig00hst}, overlaid on HST image of 
the galaxy \citep{caldwell1999}. The D100 tail is also bright in soft 
X-rays (see Fig.~\ref{Fig00hst}, left panel). {\it GALEX} observations 
further revealed a UV component of the tail in the inner $\sim 15$~kpc 
of its length \citep{smith2010}. Some star formation is thus likely 
taking place in the stripped material. No \ion{H}{1} was detected with 
VLA in the galaxy or the tail \citep{bravoalfaro2000, bravoalfaro2001}. 

D100 is a $0.3~L_*$ SBab type galaxy with estimated stellar mass of 
$2.1\times 10^9~M_\odot$ \citep{yagi2010}\footnote{The stellar mass 
estimate comes from the MPA-JHU SDSS catalog in the DR7\_v5.2 
version.}. It exhibits starburst in its core, with current star 
formation rate of $\sim 2.3~M_\odot$\,yr$^{-1}$ (derived from WISE 
band~4), and post-starburst characteristics in the rest of the disk. 
D100 is projected at only $\sim 240$~kpc from the Coma cluster center. 
Fig.~\ref{Fig00hst} (left panel) shows its position in a {\it Chandra} 
view of the central parts of the Coma cluster. Its radial velocity 
component relative to the Coma mean is $\sim -1570$~km\,s$^{-1}$ 
\citep{yagi2007}. 

In the HST image in Fig.~\ref{Fig00hst} (right panel), prominent dust 
features obscuring the eastern side of the disk are clearly visible 
\citep{caldwell1999}. They coincide with the tail direction. The image 
also reveals a strong two-armed spiral pattern extending out to $\sim 
1.4$~kpc radius. While two early-type galaxies are projected close to 
D100, GMP~2897 (D99) at $\sim 10$~kpc and GMP~2852 at $\sim 30$~kpc, 
their radial velocities are substantially larger (by about 4500 and 
2000~km\,s$^{-1}$, respectively). Another galaxy, a low-surface 
brightness GMP~2913 occurs at a projected distance of $\sim 9$~kpc from 
D100 (see Fig.~\ref{Fig00hst}, middle panel). Its radial velocity is 
only by $\sim 130$~km\,s$^{-1}$ lower than that of D100, which makes 
their interaction possible \citep{yagi2007}. However, the optical 
isophotes of D100 are symmetric and do not indicated any (strong) 
recent tidal interaction. 

In this paper we report our discovery of abundant molecular gas 
component in the prominent tail of D100. We study its distribution and 
kinematics and compare it with the warm ionized component, as well as 
discuss the origin of the tail and of the molecular gas in the tail. 
The paper is organized as follows: our observations and results are 
introduced in Sections~2 and 3, kinematics of the tail is studied in 
Section~4, and the distribution of the CO emission and its correlation 
with H$\alpha$ emission in Section~5. Also the CO-to-H$_2$ conversion 
factor is discussed in Section~5. In Section~6, we discuss the origin 
of the tail, the efficiency of star formation in the stripped gas, and 
we study the orbit of the galaxy in Coma. Conclusions are drawn in 
Section~7. 
Throughout the paper, we use the Coma distance of $97.5$~Mpc, thus 
$1~\rm{arcsec}\approx 0.473$~kpc. Cosmological constants in use were 
$H_0= 70$, $\Omega_\Lambda= 0.73$, $\Omega_m= 0.27$.

\section{Observations}\label{observations}
The observations were carried out with the IRAM 30m antenna operated by 
the Institut de Radio Astronomie Millim\'etrique (IRAM) at Pico Veleta, 
Spain, in December 2014. The EMIR receiver in E090 and E230 bands was 
used to observe simultaneously at the frequencies of the 
$^{12}$CO(1--0) ($\nu_{\rm rest}= 115.271$~GHz) and the $^{12}$CO(2--1) 
($\nu_{\rm rest}= 230.538$~GHz) lines. Observing conditions were 
excellent with PWV as low as $1-2$~mm and system temperatures typically 
about $140-180$~K at CO(1-0) and $160-200$~K at CO(2-1). The FTS 
spectrometer with $\sim 200$~kHz spectral resolution was connected to 
both lines. Also the WILMA autocorrelator with a spectral resolution of 
2~MHz at both 115~GHz and 230~GHz was used as a back-up. The 
observations were done in a symmetric Wobbler switching mode with the 
maximum throw of the secondary reflector of $240''$ in order to avoid 
with OFF positions the tail if oriented in azimuth.

The half power beamwidth (HPBW) of the IRAM 30m main beam is described 
to a good accuracy by HPBW$('')= 2460/\nu(\rm{GH}z)$. Thus at the 
CO(1-0) and CO(2-1) sky frequencies, HPBW$\sim 21.7''\approx 10.3$~kpc 
and $10.9''\approx 5.2$~kpc, respectively. The corresponding main beam 
projected area $\Omega_B \simeq 533$~arcsec$^2= 120$~kpc$^2$ and 
$\simeq 135$~arcsec$^2= 31$~kpc$^2$, respectively, including a factor 
$1/\ln2$ of a Gaussian beamshape correction. 

\begin{figure*}[t]
\centering
\includegraphics[width=0.98\textwidth]{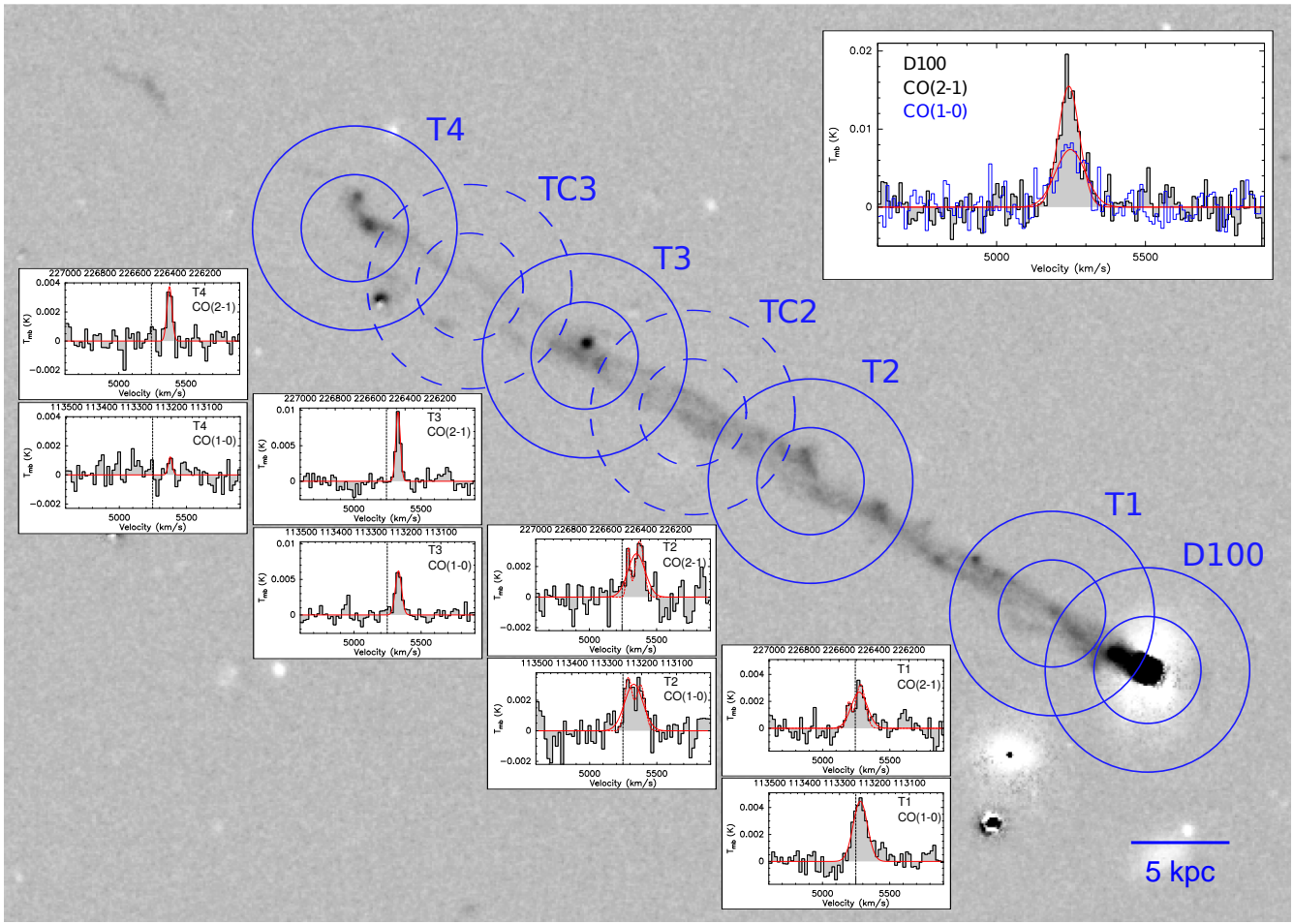}
\caption{
The CO(1-0) and CO(2-1) spectra measured in D100 (top right panel, 
smoothed to 10.3~km\,s$^{-1}$ channels) and the tail regions T1$-$T4 
(smoothed to 20~km\,s$^{-1}$). The IRAM 30m $^{12}$CO(1-0) and 
$^{12}$CO(2-1) beams (HPBW of $21.7''\approx 10.3$~kpc and 
$10.9''\approx 5.2$~kpc, respectively) are shown for the observed 
pointings overlaid on optical H$\alpha$ image of D100 \citep[Subaru 
Telescope,][]{yagi2010}. Parameters of the CO lines Gaussian fits (red 
curves) are given in Table~\ref{TabRes}. Dashed vertical lines in the 
tail spectra indicate the central velocity of the CO(1-0) line in the 
main body. The strongest CO emission in the tail is clearly in the 
region T3, at about $1'.1\approx 32$~kpc distance from the galaxy. The 
spectra from the two complementary regions TC2 and TC3 are shown in 
Fig.~\ref{Fig003B}. The CO velocity scale is LSR; for the sky position 
of D100, $V_{\rm hel}= V_{\rm LSR} - 8.5$~km\,s$^{-1}$. 
}\label{Fig003}
\end{figure*}

Seven integration points were selected to cover the main body of D100, 
as well as most of its H$\alpha$ bright tail, out to a projected 
distance of $\sim 1'.6\approx 45$~kpc from the galaxy (see the scheme 
in Fig.~\ref{Fig003}). During the observation we first focused on 
H$\alpha$ bright regions in the tail (pointings T1--T4) and only after 
revealing strong CO emission we moved in the remaining time also to 
intermediate parts of the tail with less (or less clumpy) H$\alpha$ 
emission ('complementary' pointings TC2 and TC3). The list of observed 
positions is given in Tab.~\ref{Sources}, together with information on 
actual on-source observing times. 

The FTS backend often suffered from ``platforming'' between individual 
units. There is a correction CLASS script 
FtsPlatformingCorrection5.class\footnote{
www.iram.es/IRAMES/mainWiki/AstronomerOfDutyChecklist} that can 
subtract baselines individually from the affected FTS units. However, 
our expected lines were placed in the middle of the central FTS unit, 
thus platforming was not a big issue. 

\begin{table}
\centering
\caption[]{List of observed positions.}
\label{Sources}
\begin{tabular}{lccccc}
\hline
\hline
\noalign{\smallskip}
&R.A.& Dec. & $d_{\rm D100}$ & $T_{\rm ON}$\\
&(J2000) & (J2000) & (kpc) & (min)\\
\noalign{\smallskip}
\hline
\noalign{\smallskip}
D100 & 13:00:09.14 & +27:51:59.2 &    - &  99\\
T1   & 13:00:09.88 & +27:52:04.1 &  5.2 & 196\\ %11'' 
T2   & 13:00:11.84 & +27:52:19.4 & 19.4 & 112\\ %41.1'' D100-T4
TC2  & 13:00:12.77 & +27:52:26.2 & 26.1 &  83\\ %55.2'' D100-T6
T3   & 13:00:13.64 & +27:52:32.4 & 32.3 & 127\\ %68.3'' D100-T2
TC3  & 13:00:14.53 & +27:52:39.0 & 38.7 & 106\\ %81.8'' D100-T5
T4   & 13:00:15.45 & +27:52:45.3 & 45.2 & 177\\ %95.53''D100-T3
\noalign{\smallskip}
\hline
\noalign{\smallskip}
\end{tabular}
\end{table}

The data were reduced in the standard manner using CLASS from the 
GILDAS\footnote{http://www.iram.fr/IRAMFR/GILDAS} software package 
developed at IRAM. Bad scans were flagged and emission line-free 
channels in the total width of about 1000~km\,s$^{-1}$ were used to 
subtract (mostly) first-order baselines. The corrected antenna 
temperatures, $T^*_{\rm A}$, provided by the IRAM 30m calibration 
pipeline, were converted to main-beam brightness temperature by $T_{\rm 
mb}=T^*_{\rm A} F_{\rm eff}/ \eta_{\rm mb}$, using a main beam 
efficiency of about $\eta_{\rm mb}= 0.78$ at 115~GHz and $0.59$ at 
230~GHz, and the forward efficiencies $F_{\rm eff}$ of 0.94 and 0.92, 
respectively. The rms noise levels typically of $1-1.5$~mK per 
10.6~km~s$^{-1}$ channels were obtained. Gaussian fits were used to 
measure peak $T_{\rm mb}$, width, and position of the detected CO 
lines. The flux density-to-antenna temperature conversion factor is 
$S_\nu/ T_{\rm A}^*= 3514/ (\eta_a D^2)$, where $\eta_a$ is the 
telescope aperture efficiency and $D$ is the diameter of the telescope 
in meters. The IRAM 30m aperture efficiency is $\sim 0.6$ at CO(1-0) 
frequency and $\sim 0.41$ at CO(2-1) frequency. The $S_\nu/T_{mb}$ 
conversion is thus $\sim 5$~Jy\,beam$^{-1}$\,K$^{-1}$ for both bands.

\section{Results}\label{results}

\subsection{Main body of D100}
The CO(1-0) and CO(2-1) spectra measured in the main body D100 pointing 
are shown in Fig.~\ref{Fig003} (top-right panel). Both lines are 
clearly detected, the CO(2-1) line is about two-times stronger than 
CO(1-0). We calculate the CO luminosity from the standard relation of 
\citet{solomon2005}
\begin{equation}\label{luminosity}
L'_{\rm CO}= 3.25\times 10^7\, S_{\rm CO}\, \Delta\upsilon\, \nu_{\rm 
obs}^{-2}\, D_L^2\, (1+z)^{-3}, 
\end{equation}
where $L'_{\rm CO}$ is the CO line luminosity in $\rm K\, km\, s^{-1}\, 
pc^2$, $S_{\rm CO}\, \Delta\upsilon$ is the CO velocity integrated line 
flux in $\rm Jy\, km\, s^{-1}$, $\nu_{\rm obs}$ is the observed CO line 
frequency in GHz, and $D_L$ is the distance in Mpc. The CO(1-0) 
luminosity is $\sim 1.1\times 10^8$~$\rm K\, km\, s^{-1}\, pc^2$. 
Following 
\begin{equation}\label{mass}
M_{\rm H_2}\, [M_\odot]= 4.5\, L'_{\rm CO}\, [{\rm 
K\,km\,s^{-1}\,pc^2}], 
\end{equation}
where we assume a CO/H$_2$ conversion factor of $2\times 
10^{20}$~cm$^{-2}$(K\,km\,s$^{-1})^{-1}$ that is standard under Milky 
Way disk conditions \citep[e.g.,][]{bolatto2013}. However, the actual 
value of the conversion factor is uncertain, especially in the tail of 
D100. This is discussed later in Section~\ref{x-factor}. The luminosity 
corresponds to a molecular gas mass of $\sim 4.8\times 10^8 M_\odot$ 
(including a factor of 1.36 to account for Helium). 

The CO lines are in Fig.~\ref{Fig003} fitted with Gaussians. Their 
parameters are given in Table~\ref{TabRes}. The CO(1-0) central 
velocity is $\sim 5248$~km\,s$^{-1}$, assuming the radio definition of 
the velocity ($\upsilon_{\rm rad}/c= 1- \nu_{\rm sky}/ \nu_{\rm 0}= 
z/(z+1)$). This is consistent with the optical radial velocity 
$\upsilon_{\rm opt}= c\,(\nu_{\rm 0}/ \nu_{\rm sky} -1)= c\,z= 
5348$~km\,s$^{-1}$ measured from H$\alpha$, [NII], and [SII] emission 
in the galaxy center \citep{yagi2007}. The CO(1-0) line width is $\sim 
120$~km\,s$^{-1}$, contrary to $\sim 80$~km\,s$^{-1}$ for the CO(2-1) 
line. The difference may reflect detecting emission from larger 
galactocentric radii with increasing rotation velocity within the 
larger CO(1-0) beam. From the Tully-Fisher relation 
\citep[e.g.,][]{gnedin2007}, a disk galaxy with D100's stellar mass has 
a typical circular velocity of $\sim 80$~km\,s$^{-1}$. Given the 
inclination of the galaxy of $\sim 43^\circ$ (see 
Section~\ref{SecOrbit}), this corresponds to the linewidth of $2\times 
v_{\rm circ}\,\sin{43^\circ}\approx 110$~km\,s$^{-1}$, thus consistent 
with the measured CO(1-0) linewidth. The larger CO(1-0) beam is also 
likely contaminated by the emission from the inner tail (see lower).

The CO (2-1)/(1-0) integrated luminosity ratio measured in the D100 
pointing is $\sim 1.5$ (and the ratio of the peak temperatures is $\sim 
2.2$), not corrected for the different beam sizes. Assuming a typical 
CO (2-1)/(1-0) ratio of $0.8- 1$, the measured value is consistent with 
emission coming from an extended source that is larger than the CO(2-1) 
beam.

\subsection{Molecular gas in the tail of D100}

\begin{table*}[t]
\centering
\caption[]
{Properties of the detected CO lines in D100.}
\label{TabRes}
\begin{tabular}{llcccccccc}
\hline
\hline
\noalign{\smallskip}
Source & Line & rms & velocity & rel. vel. & FWHM & $T_{\rm peak}$ & $I_{\rm CO, fit}$ & $L_{\rm CO}$ & $M_{\rm mol}$\tablenotemark{a}\\
     &        & (mK) & (km\,s$^{-1}$) & (km\,s$^{-1}$) & (km\,s$^{-1}$) & (mK) & (K\,km\,s$^{-1}$) & ($10^7$~K\,km\,s$^{-1}$pc$^2$) & ($10^8 M_\odot$)\\
\noalign{\smallskip}
\hline
\noalign{\smallskip}
D100 & CO(1-0) & 1.7 & $5247.9\pm  5.6$ & 0 & $118.0\pm 14.2$ &  7.4 & $0.93\pm 0.09$ & $10.62\pm 1.03$ & $4.78\pm 0.46$\\
     & CO(2-1) & 1.7 & $5242.7\pm  2.5$ & 0 & $ 83.9\pm  6.5$ & 16.  & $1.38\pm 0.09$ & $3.94\pm 0.26$ & $2.17\pm 0.14$\\
\noalign{\smallskip}\hline\noalign{\smallskip}
T1   & CO(1-0) & 1.0 & $5280.7\pm  5.6$ & 33 & $122.9\pm 13.7$ &  4.5 & $0.59\pm 0.05$ & $6.74\pm 0.57$ & $3.03\pm 0.26$\\
\noalign{\smallskip}
     & CO(2-1) & 1.1 & $5272.4\pm 10.0$ & 30 & $145.5\pm 25.5$ &  2.7 & $0.43\pm 0.06$ & $1.23\pm 0.14$ & $0.68\pm 0.08$\\
     & CO(2-1) \#1&     & $5193.2\pm  9.1$ & -55 & $ 41.6\pm 19.7$ &  1.9 & $0.08\pm 0.04$ & $0.23\pm 0.11$ & $0.13\pm 0.06$\\
     & CO(2-1) \#2&     & $5282.0\pm  6.8$ & 39 & $ 82.5\pm 25.9$ &  3.3 & $0.29\pm 0.06$ & $0.83\pm 0.14$ & $0.46\pm 0.08$\\
\noalign{\smallskip}\hline\noalign{\smallskip}
T2   & CO(1-0) & 1.3 & $5328.7\pm 12.4$ & 81 & $165.3\pm 27.4$ &  3.1 & $0.55\pm 0.08$ & $6.28\pm 0.91$ & $2.82\pm 0.41$\\
     & CO(1-0) \#1&     & $5287.4\pm  9.5$ & 40 & $ 66.0\pm 19.5$ &  3.4 & $0.24\pm 0.07$ & $2.74\pm 0.80$ & $1.23\pm 0.36$\\
     & CO(1-0) \#2&     & $5379.1\pm 11.6$ & 131 & $ 77.2\pm 25.7$ &  3.0 & $0.25\pm 0.08$ & $2.86\pm 0.91$ & $1.29\pm 0.41$\\
\noalign{\smallskip}
     & CO(2-1) & 1.3 & $5350.0\pm 13.8$ & 107 & $151.1\pm 30.4$ &  2.9 & $0.46\pm 0.08$ & $1.31\pm 0.23$ & $0.72\pm 0.13$\\
     & CO(2-1) \#1&     & $5289.4\pm  6.3$ & 47 & $ 33.3\pm 15.6$ &  3.1 & $0.11\pm 0.04$ & $0.31\pm 0.11$ & $0.17\pm 0.06$\\
     & CO(2-1) \#2&     & $5374.5\pm  7.1$ & 132 & $ 72.2\pm 17.6$ &  3.7 & $0.29\pm 0.06$ & $0.83\pm 0.14$ & $0.46\pm 0.08$\\
\noalign{\smallskip}\hline\noalign{\smallskip}
T3   & CO(1-0) & 1.3 & $5331.3\pm  3.0$ & 83 & $ 52.8\pm  7.4$ &  6.4 & $0.36\pm 0.04$ & $4.11\pm 0.46$ & $1.85\pm 0.21$\\
     & CO(2-1) & 1.3 & $5328.4\pm  1.7$ & 86 & $ 37.7\pm  4.3$ & 10.7 & $0.43\pm 0.04$ & $1.23\pm 0.11$ & $0.68\pm 0.06$\\
\noalign{\smallskip}\hline\noalign{\smallskip}
T4   & CO(1-0) & 1.0 & $5379.5\pm  8.6$ & 132 & $ 31.3\pm 18.5$ &  1.4 & $0.05\pm 0.03$ & $0.57\pm 0.34$ & $0.26\pm 0.15$\\
     & CO(2-1) & 1.0 & $5377.5\pm  3.6$ & 135 & $ 44.6\pm  8.1$ &  3.7 & $0.18\pm 0.03$ & $0.51\pm 0.09$ & $0.28\pm 0.05$\\
\noalign{\smallskip}\hline\hline\noalign{\smallskip}
TC2  & CO(1-0) & 1.5 & - & - & - & - & - & - & - \\
     & CO(2-1) & 1.3 & - & - & - & - & - & - & - \\
\noalign{\smallskip}\hline\noalign{\smallskip}
TC3  & CO(1-0) & 1.2 & $5350.7\pm 10.6$ & 103 & $ 91.1\pm 26.4$ &  2.8 & $0.27\pm 0.06$ & $3.08\pm 0.69$ & $1.39\pm 0.31$\\
     & CO(2-1) & 1.5 & - & - & - & - & - & - & - \\
\noalign{\smallskip}\hline\noalign{\smallskip}
\end{tabular}
\tablenotetext{1}{There is an additional systematic uncertainty of 
a factor of several due to CO-to-H$_2$ conversion relation.}
\tablecomments{
The table gives the $1\sigma$ rms in 10.2~km\,s$^{-1}$ channels, 
parameters of (multiple) Gaussian fits (the line heliocentric central 
velocity, the FWHM, the peak temperature, and the integrated 
intensity), the measured integrated intensity, and the molecular gas 
mass. The temperatures are given in $T_{\rm mb}$ scale. Baselines were 
subtracted in the velocity range $4200 - 6200$~km\,s$^{-1}$. Radio 
definition of velocity was used to convert the sky frequency of the 
source. 
}
\end{table*}

Our new CO observations of the D100 tail reveal for the first time the 
presence of abundant molecular gas coexisting with the previously 
observed H$\alpha$ and X-ray components. Figure~\ref{Fig003} depicts 
CO(1-0) and CO(2-1) spectra measured in the T1$-$T4 pointings. The 
complementary spectra from the two less sensitive regions TC2 and TC3 
are shown in Fig.~\ref{Fig003B}. The parameters of the Gaussian fits to 
the detected lines are summarized in Table~\ref{TabRes}. Further in 
this section, we divide the tail into three parts (inner, intermediate, 
and outer). As the morphology of the tail does not change substantially 
along its length, we base the division only on the projected distance 
from the galaxy. 

As noted above, the CO/H$_2$ conversion relation is uncertain in the 
tail of D100. It may introduce some systematic errors into the values 
of H$_2$ masses derived in the following subsections from the observed 
CO(1-0) luminosities. While it is hard to quantify the uncertainty in 
the conversion factor, it may be a factor of a few, but likely less 
than an order of magnitude. For more details see the discussion in 
Sec.~\ref{x-factor}.

\subsubsection{Inner tail}
In the innermost tail pointing T1, CO(1-0) emission is strong and 
comparable to that coming from the D100 disk itself. Following 
Eqs.~\ref{luminosity} and \ref{mass}, the corresponding H$_2$ mass is 
$\sim 3\times 10^8~M_\odot$. CO(2-1) emission is substantially weaker 
than in the disk -- the ratio of integrated intensities is low, $I_{\rm 
CO(2-1)}/I_{\rm CO(1-0))}\sim 0.7$. This is very likely due to 
contribution from the disk covered by the outer parts of the larger 
CO(1-0) beam. The values of the CO~(2-1)/(1-0) line-ratios are for the 
observed regions depicted in Fig.~\ref{FigLineratios}. 

As compared to the main body pointing, the spectral lines are in T1 
shifted by $\sim 40$~km\,s$^{-1}$ to higher velocities. The CO(2-1) 
line profile moreover suggests the emission is coming from two 
substructures at different radial velocities, with a separation $\Delta 
\upsilon \approx 90$~km\,s$^{-1}$. We fit the substructures with a 
double Gaussian (see Table~\ref{TabRes}). The lower-velocity peak is at 
$\sim 50$~km/s below the D100 central velocity and the higher-velocity 
peak at $\sim 40$~km/s above it (however, the latter component is much 
stronger than the former one). This could be an imprint of the galactic 
rotation.

\subsubsection{Intermediate tail}
At the projected distance of $\sim 20$~kpc from the galaxy, in the 
second closest region T2, both CO(1-0) and CO(2-1) emission are again 
bright. The CO intensity corresponds to H$_2$ mass of $\sim 2.8\times 
10^8~M_\odot$. The linewidths of $\sim 150$~km\,s$^{-1}$ are larger 
than in the main body and in the T1 region, but the line profiles 
clearly suggest there are (at least) two distinct peaks with $\Delta 
\upsilon \approx 90$~km\,s$^{-1}$. Presumably, the two peaks could 
correspond to the substructures (bifurcation) weakly visible in the 
H$\alpha$ image. The FWHMs of the T1 and T2 sub-peaks are $\sim 
30-80$~km\,s$^{-1}$, which corresponds to the velocity dispersion range 
of $\sim 13-34$~km\,s$^{-1}$. Much better spatial and spectral 
resolution are needed to further resolve smaller entities, such as 
individual molecular clouds whose velocity dispersions are smaller, 
typically $\sim 10$~km\,s$^{-1}$. 

Assuming a NFW profile for the Coma mass distribution 
\citep{navarro1996, kubo2007}, and the DM halo mass of D100 of $\sim 
2\times 10^{11}~M_\odot$ \citep[following the relation 
of][]{behroozi2010}, we estimate the tidal truncation radius 
$R_t\approx R(M_{\rm D100}/2\times M_{\rm Coma})^{1/3}$ at the 
projected galactocentric distance $R= 240$~kpc of D100 in Coma to be 
$\lesssim 20$~kpc. Thus, the material occurring in the pointing T2 (and 
farther out) is beyond the tidal truncation radius and will likely 
contribute to the intra-cluster medium. However, this does not take 
into account the gas velocities relative to the escape speed from the 
galaxy's potential. Thus, even gas inside the tidal truncation will 
likely join the ICM since it will exceed the escape speed due to ram 
pressure acceleration.

The strongest emission (in terms of the S/N ratio) comes from the $\sim 
30$~kpc distant region T3. The corresponding molecular gas mass is 
$\sim 1.9\times 10^8~M_\odot$. The lines are substantially narrower 
than in the inner tail - the linewidth drops to $\sim 40- 
50$~km\,s$^{-1}$, with no substructures suggested. This could be due to 
leaving behind the circular velocity component and/or due to a compact 
size of the CO-emitting region, possibly associated with the bright 
compact H$\alpha$ (probably \ion{H}{2}) region clearly visible in the 
center of the T3 beam. 

The CO (2-1)/(1-0) integrated intensity line ratios (not corrected for 
the beam sizes) are in the T2 and T3 regions again rather low, $\sim 
0.8$ and $\sim 1.2$, respectively (see Fig.~\ref{FigLineratios}). This 
is probably due to more CO emission coming from the outer parts of the 
larger CO(1-0) beam, outside the CO(2-1) beam. However, taking the 
distribution of H$\alpha$ emission as a proxy for the CO distribution, 
most of the emission may be in the T3 region enclosed by the smaller 
CO(2-1) beam. This would then suggest a low (corrected) line ratio. In 
the perpendicular direction, we assume the distribution of the CO 
emission is confined to the width of the H$\alpha$ (and X-ray) tail.

\subsubsection{Outer tail}
\begin{figure}[t]
\centering
\includegraphics[height=0.45\textwidth,angle=270]{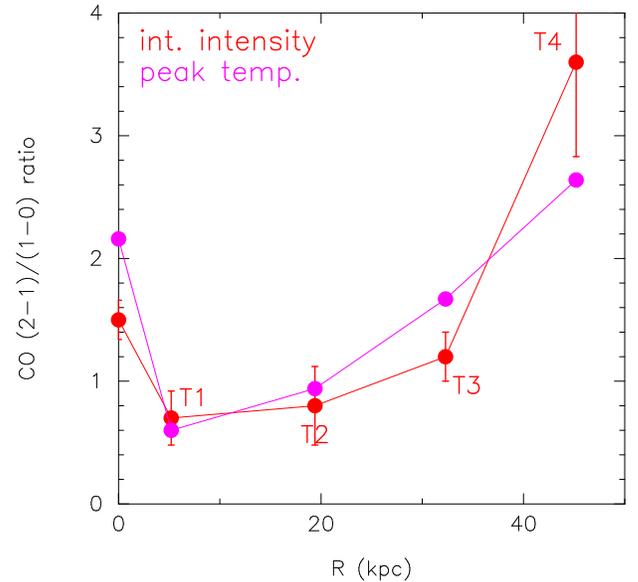}
\caption{
CO (2-1)/(1-0) integrated intensity and peak temperature line ratios 
measured in D100 and the tail regions T1 to T4. The values are not 
corrected for different beam sizes. With four times the beam area, a 
compact source will experience four times the beam dilution in the 
CO(1-0) beam. 
}\label{FigLineratios}
\end{figure}

In the most distant observed pointing T4, at $\sim 45$~kpc from the 
galaxy, CO(2-1) emission is reliably detected, while there is only a 
hint of CO(1-0) at the same velocity. The line intensity corresponds to 
H$_2$ mass of $\sim 2.6\times 10^7~M_\odot$. The CO(2-1) line is again 
narrow (${\rm FWHM}\sim 40$~km\,s$^{-1}$) and shifted towards higher 
velocities, by about 150~km\,s$^{-1}$ relative to the main body. 
Contrary to the previous pointings, the integrated intensity line ratio 
in T4 is $\sim 3.6$, thus consistent with a point-source emission 
diluted by the four-times larger CO(1-0) beam (and a typical R$_{21}$ 
line ratio of $\sim 0.8-1$). Alternatively, it could suggest that 
molecular gas is in T4 more extended, and heated to higher 
temperatures, indicating optically thin gas \citep[see 
e.g.,][]{crosthwaite2007}. Fig.~\ref{FigLineratios} clearly illustrates 
that along the tail the integrated intensity line ratio (not corrected 
for the different beam sizes) increases.

\begin{figure}[t]
\centering
\includegraphics[height=0.49\textwidth,angle=270]{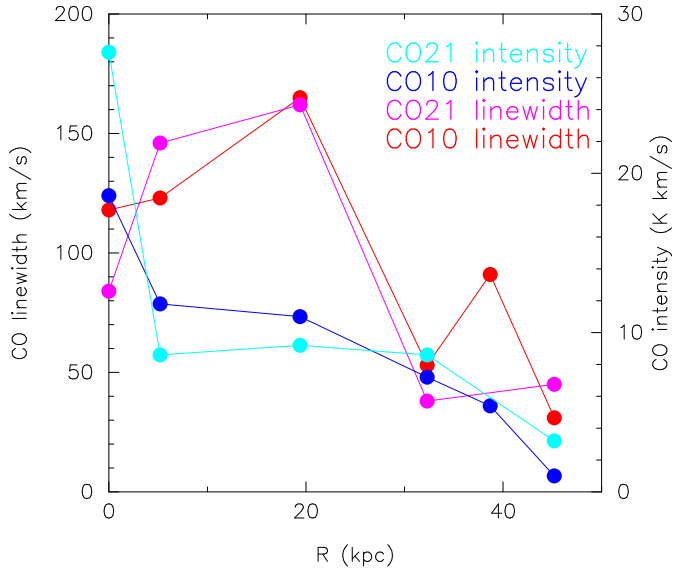}
\caption{
CO(1-0) and CO(2-1) peak temperatures and integrated intensities in 
observed positions along the tail as a function of projected distance 
from D100.
}\label{Fig006}
\end{figure}

In Fig.~\ref{Fig006}, the peak CO main beam temperatures are shown as s 
function of the downstream distance in the tail, together with the CO 
integrated intensities. 

\subsubsection{Complementary, less-sensitive pointings}

\begin{figure}[t]
\centering
\includegraphics[width=0.22\textwidth]{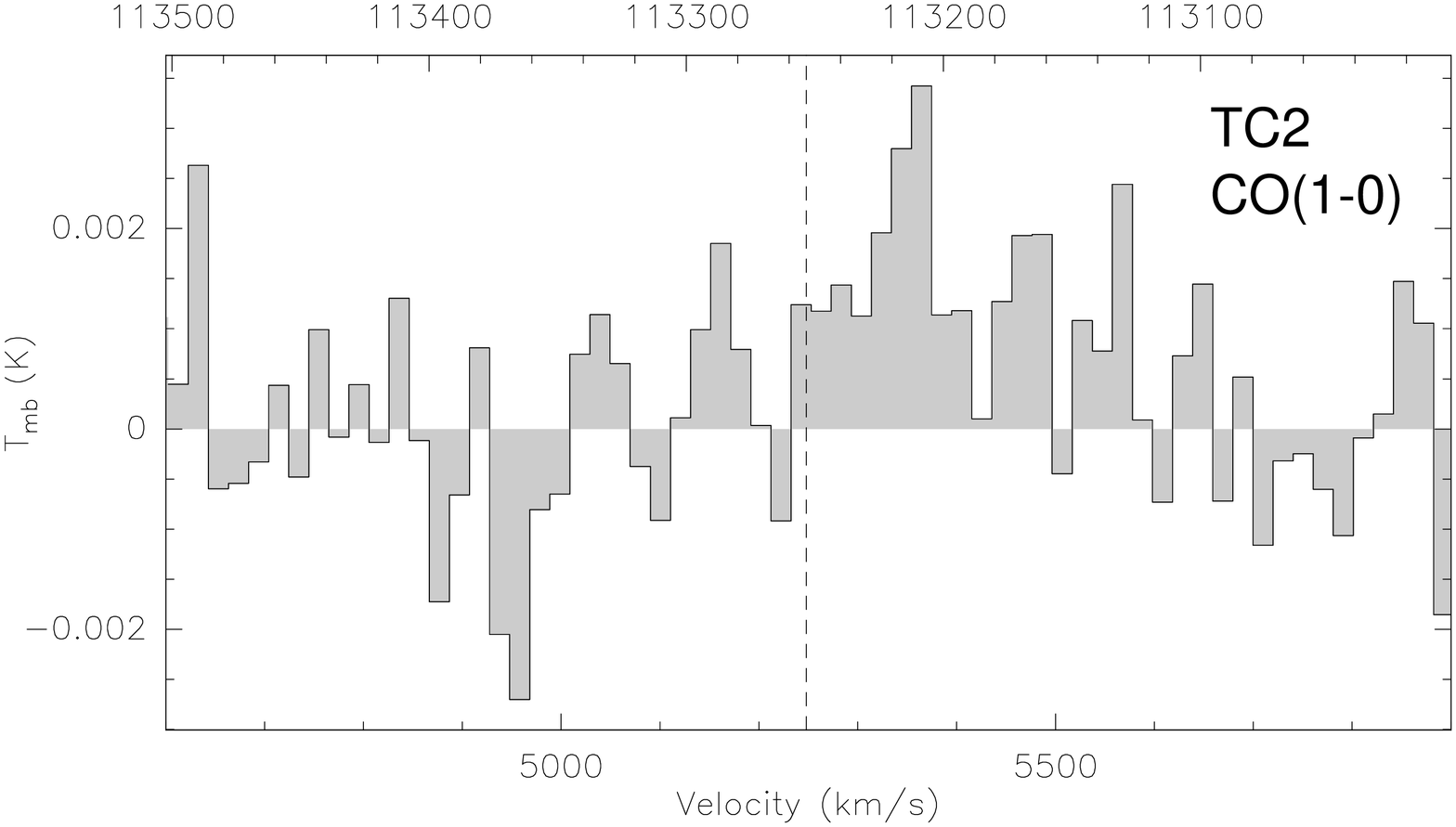}
\hspace{0.1cm}
\includegraphics[width=0.22\textwidth]{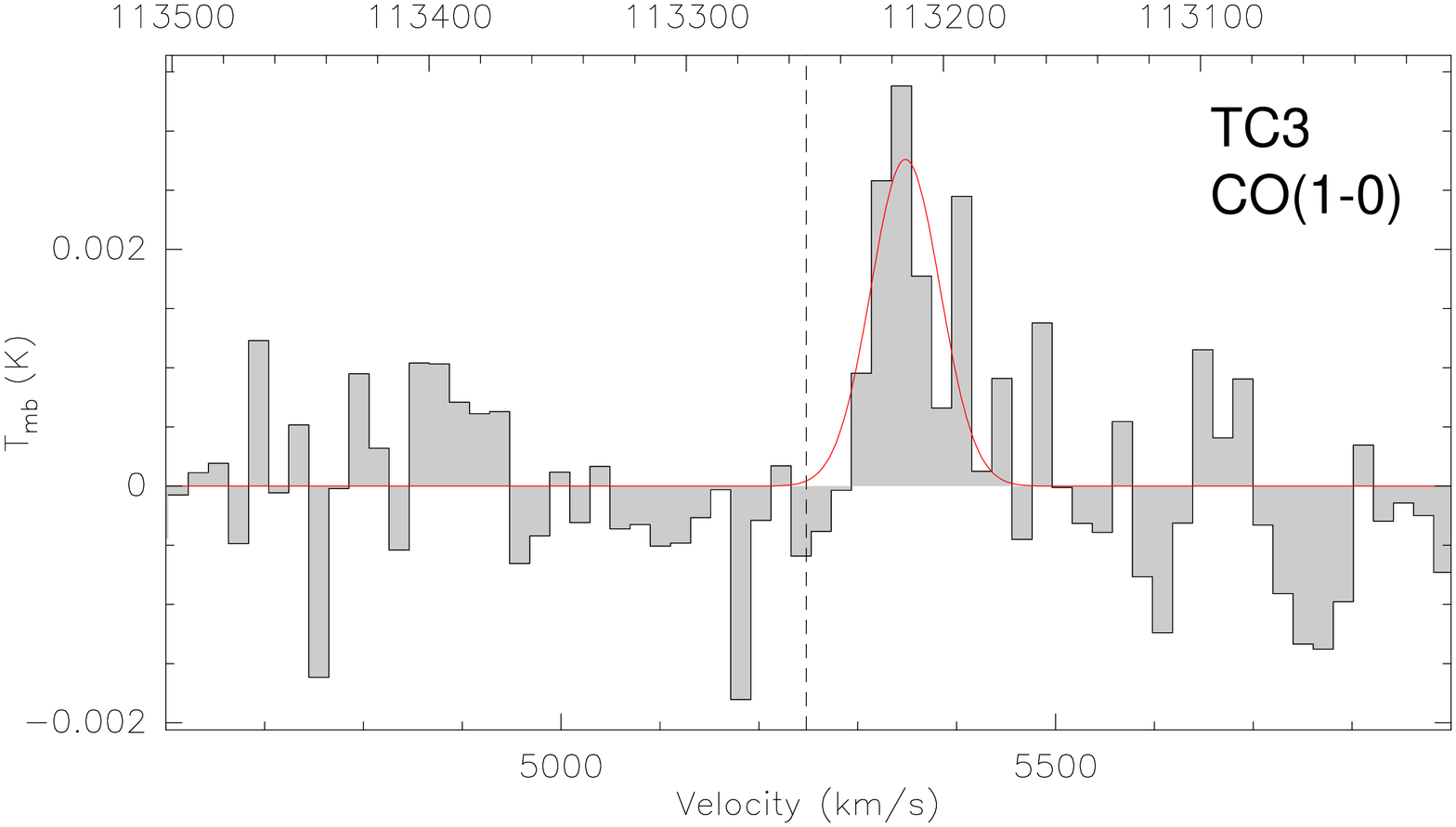}\\
\vspace{0.1cm}
\includegraphics[width=0.22\textwidth]{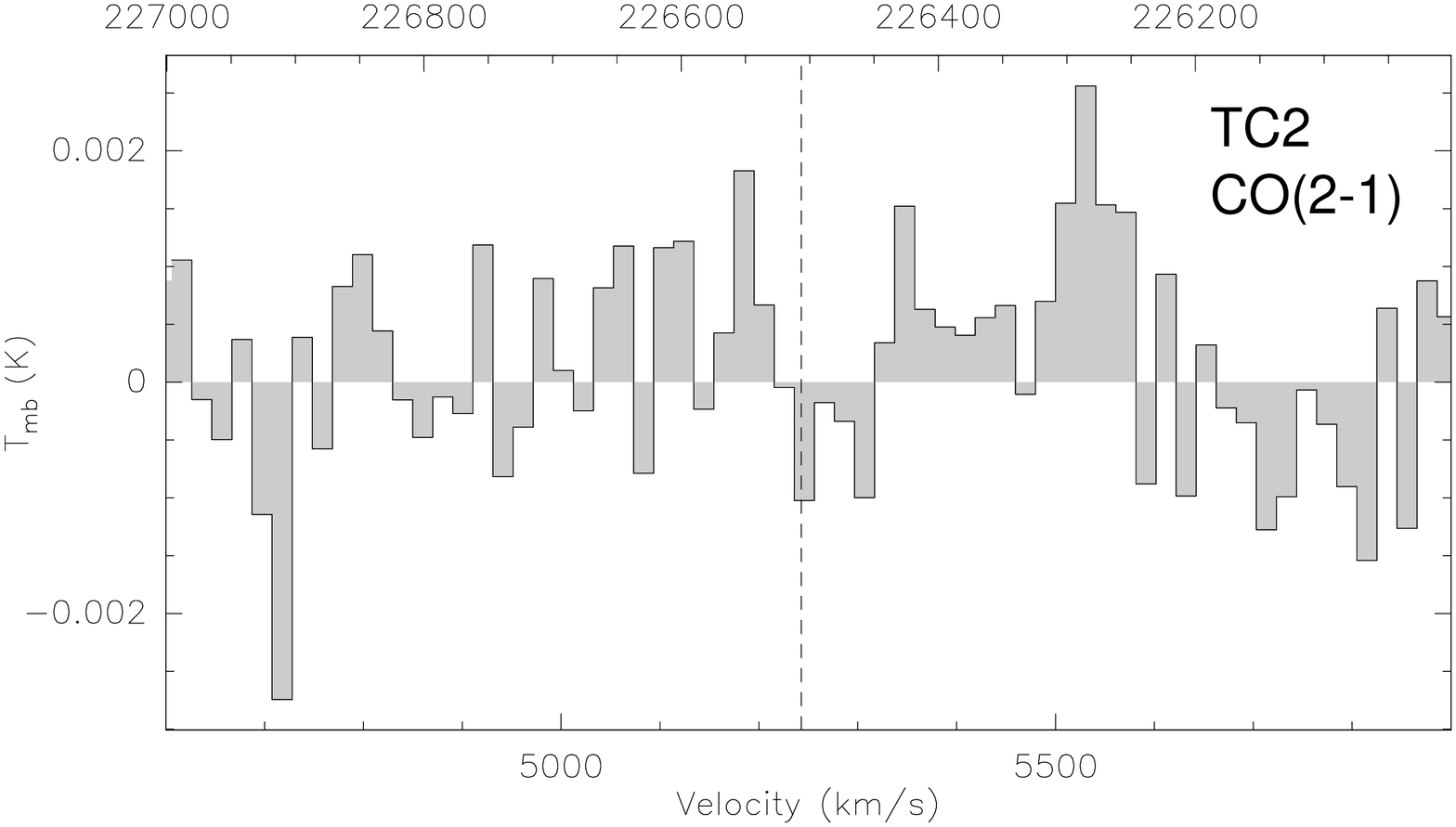}
\hspace{0.1cm}
\includegraphics[width=0.22\textwidth]{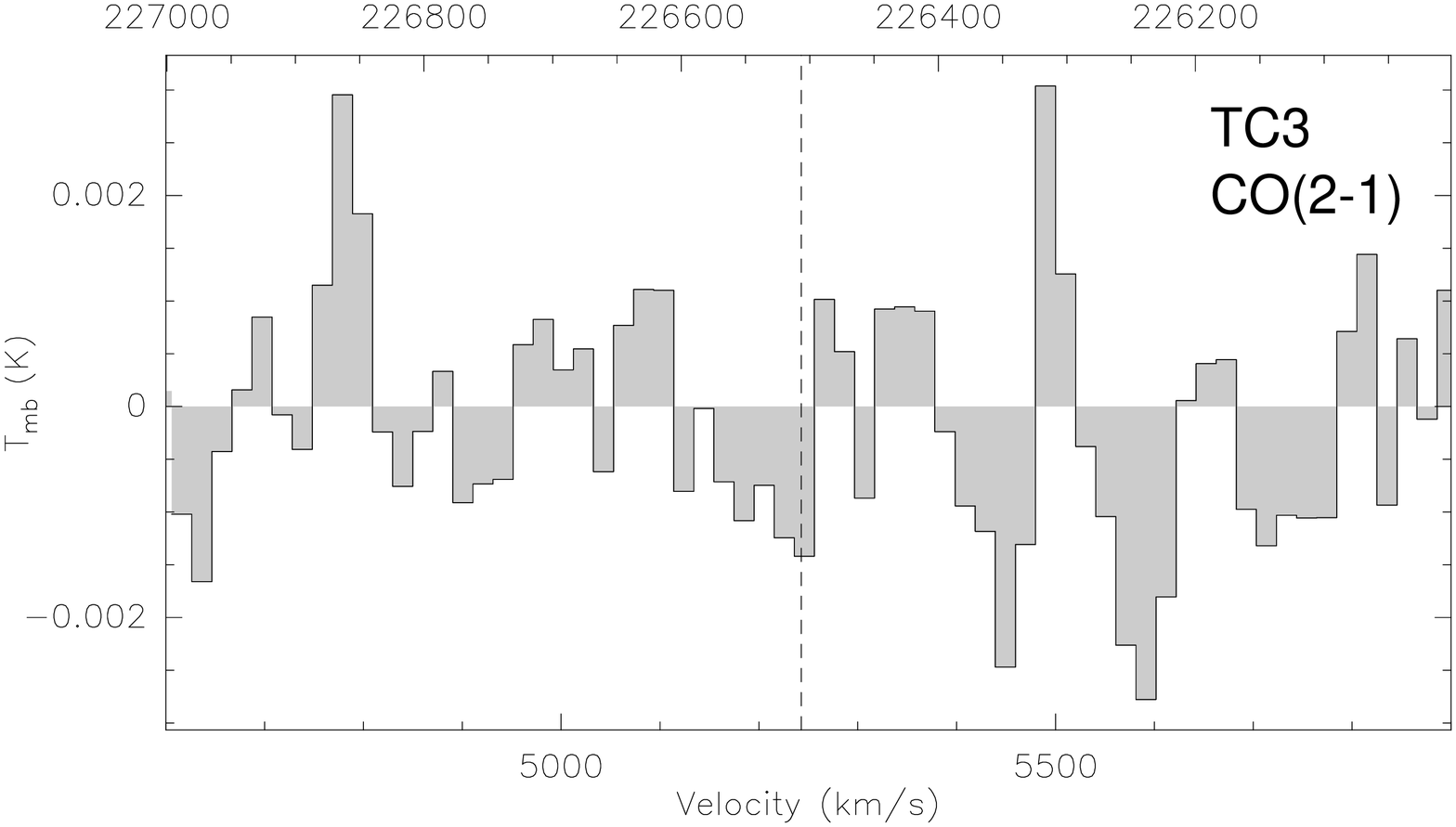}
\caption{
CO(1-0) and CO(2-1) spectra measured in the complementary regions TC2 
and TC3 in the D100 tail (see the scheme in Fig.~\ref{Fig003}). The rms 
sensitivity in CO(1-0) and CO(2-1) is 1.5~mK and 1.3~mK, respectively, 
for the TC2 region, and 1.2~mK and 1.5~mK, respectively, for the TC3 
region. Dashed vertical lines indicate the central velocity of the 
CO(1-0) line in the main body spectrum.
}\label{Fig003B}
\end{figure}

In addition to the H$\alpha$ bright regions T1-T4 that revealed strong 
CO emission, less sensitive observations in regions TC2 and TC3 
covering the parts of the tail with less (or less clumpy) H$\alpha$ 
emission were performed. These regions would need more observing time 
to either confirm detection or decrease rms. It is still interesting to 
inspect their spectra in Fig.~\ref{Fig003B}. They suggest that CO(1-0) 
emission is there stronger than CO(2-1), contrary to the neighboring 
regions T3 and T4. In TC2, at a projected distance of $\sim 26$~kpc 
from the galaxy (see the scheme in Fig.~\ref{Fig003}), there is a hint 
of CO(1-0) emission, with a S/N ratio of $\sim 3-4$, while CO(2-1) 
emission is absent. In TC3, at $\sim 39$~kpc, CO(1-0) emission is 
(marginally) detected with a S/N of $\sim 4.6$, but CO(2-1) is not 
detected. The corresponding H$_2$ mass in TC3 is $\sim 1.4\times 
10^8~M_\odot$. 

The presumably low CO~(2-1)/(1-0) line ratios in the TC2 and TC3 
regions may correspond to emission coming mainly from the outer parts 
of the larger CO(1-0) beams. There indeed seems to be less H$\alpha$ 
emission, especially in the location of TC3, which could also point to 
less molecular emission, assuming there is a correlation between 
H$\alpha$ and CO emission. Moreover, in both TC2 and TC3, the H$\alpha$ 
emission is probably smoother, with no obvious compact regions, as 
compared to the neighboring parts of the tail (see Fig.~\ref{Fig003}). 
The low line ratios in TC2 and TC3 could thus also suggest that 
conditions in the stripped gas are locally different: the gas may be 
more extended and have a lower temperature (since CO(1-0) usually 
traces more extended component and has a lower excitation temperature 
than CO(2-1)).

\section{Kinematic separation of dense gas component in the tail}\label{kinematic}
\begin{figure}[t]
\centering
\includegraphics[height=0.45\textwidth,angle=270]{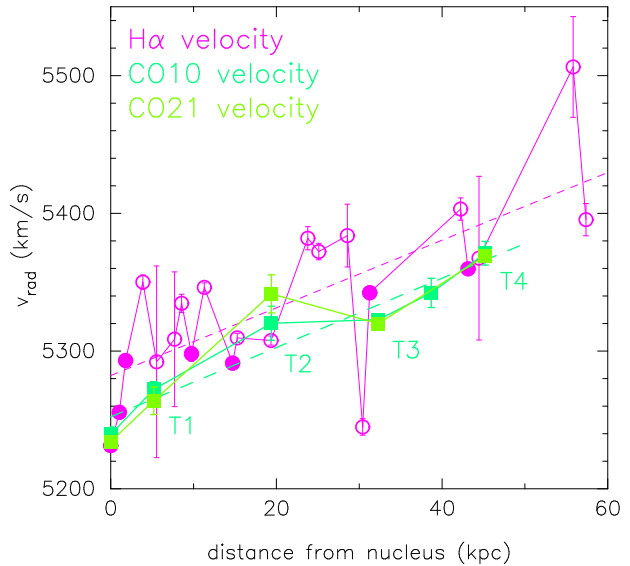}
\caption{
Velocity offset between denser, more compact CO and H$\alpha$ sources 
and more diffuse H$\alpha$ warm ionized gas along the D100 tail. Values 
derived from fitting H$\alpha$ and [NII] spectral lines \citep[revised 
from][converted to radio notation]{yagi2007}, and CO lines (single 
fits from Table~\ref{TabRes}, converted from LSR to heliocentric 
frame). Dashed lines show linear regression fits to both measurement 
sets (at distances $\gtrsim 5$~kpc). Filled circles indicate compact 
(possibly \ion{H}{2}) regions in the H$\alpha$ tail, as measured by 
H$\alpha$ surface brightness ($> 2.1\times 
10^{-17}$~erg\,s$^{-1}$\,cm$^{-2}$\,arcsec$^{-2}$). To estimate 
H$\alpha$ velocity errors, 1000 Monte Carlo simulations adding 
different noise levels to the spectra were used. Consequently, lower 
S/N measurements have larger error. 
}\label{Fig005b}
\end{figure}

Our CO observations have revealed a radial velocity gradient of $\sim 
130$~km\,s$^{-1}$ along $\sim 50$~kpc length of the D100's tail. This 
is well visible in Fig.~\ref{Fig005b} that depicts the central 
velocities of the Gaussian fits of the detected CO(1-0) and CO(2-1) 
lines as a function of their (projected) distance from the galaxy. For 
comparison, the plot also shows radial velocities measured from optical 
H$\alpha$ slit spectra in 22 regions along the D100 tail \citep[][with 
a couple of previously incorrect values revised]{yagi2007}. While 
the ionized gas velocity field is not smooth but shows small scale 
variations, a clear trend is visible: the total H$\alpha$ velocity 
difference between the galaxy center and outer tail is $\sim 
190$~km\,s$^{-1}$. It can be described by the sum of a sharp rise in 
the first 4~kpc, and a more gradual rise between $r= 4$~kpc and 
$r=60$~kpc of $\sim 140$~km\,s$^{-1}$. We note that the angular 
resolution of the two measurement sets is different -- the CO(2-1) 
beamsize is $\sim 5$~kpc, while the optical slits are typically $0.5- 
1$~kpc long.

Figure~\ref{Fig005b} reveals that radial velocities of the CO emission 
lines are systematically lower than the mean optical velocities by 
$\sim 30$~km\,s$^{-1}$, while having a similar slope. The CO velocities 
are mostly consistent with the lower edge of the velocity span of the 
H$\alpha$ regions. We identified slit regions with large H$\alpha$ 
surface brightnesses (possible \ion{H}{2} regions) and marked them in 
the plot with filled circles. All of them are below (or in one case at) 
the fit. 

This suggests that denser, more compact gas clumps, including CO and 
compact H$\alpha$ sources, are less accelerated by ram pressure due to 
their large momentum, and lag behind more diffuse H$\alpha$-emitting 
gas. Such gas lumps decouple from the surrounding diffuse gas and may 
fall back to the galaxy potential. The same phenomenon, although 
observed spatially rather than kinematically, has been observed in high 
resolution optical images of other cluster spirals, which show dense 
decoupled clouds spatially offset from more diffuse dust which is 
further downstream \citep{crowl2005, abramson2014, abramson2016, 
kenney2015}. The velocity field of the stripped ISM may have a large 
range at any distance. This was clearly shown in numerical simulations 
by \citet{tonnesen2010}. Velocity gradients have been previously 
reported in ram pressure stripped gas tails of IC~4040, IC~3418, 
NGC~4388 \citep{yoshida2012, kenney2014, oosterloo2005}. This is the 
first time that effects of differential acceleration of individual 
phases of the ISM by ram pressure are indicated from observation. 

The radial velocity gradients of $\sim 130$~km\,s$^{-1}$ in CO and 
nearly $200$~km\,s$^{-1}$ in H$\alpha$ emission form only about one 
tenth of the (projected) ICM wind speed. Simulations by 
\citet{tonnesen2012} show that even after a few hundred Myr of 
stripping, most of the gas does not reach the ICM wind velocity but has 
velocities one-third to one half of the way to it. The large momentum 
of dense molecular gas prevents ram pressure to accelerate it quickly. 
Similarly, in the RPS Virgo cluster galaxy IC~3418, \ion{H}{2} regions 
in the tail extend kinematically only to $\sim 15\%$ of the ICM wind 
velocity \citep{kenney2014}. 

Interestingly enough, the increase in velocity per unit tail length is 
constant along the tail (see the linear fits to both the H$\alpha$ and 
CO measurements in Fig.~\ref{Fig005b}). In numerical simulations, 
\citet{roediger2008} measured the velocity of the stripped gas along 
the tail to increase linearly with increasing distance from the galaxy 
(despite higher velocity width of the tail). They found that the slope 
changes for different ram pressures, i.e., it is larger for stronger 
pressures and shallower for weaker pressures (see their Figs.~12 to 
14). Also \citet{yoshida2012} measured almost monotonically increasing 
velocity of H$\alpha$ regions in the tail of IC4040 with the distance 
from the galaxy. They measured the overall acceleration rate of $\sim 
10$~km\,s$^{-1}$\,kpc$^{-1}$ in a range of $5- 80$~kpc. For comparison, 
in the tail of D100, the average acceleration is (in projection) about 
2.5~km\,s$^{-1}$\,kpc$^{-1}$. 

\begin{figure}[t]
\centering
\includegraphics[width=0.45\textwidth]{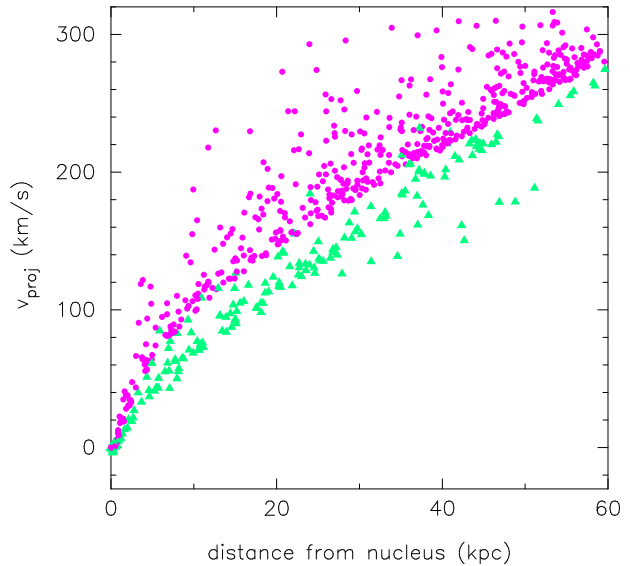}
\caption{
Effects of differential acceleration by ram pressure as shown from 
semi-analytic calculations. The plot shows distances and velocities 
relative to the disk that can reach ISM parcels with different column 
densities that are stripped from within a radius of 5~kpc from a model 
galaxy disk. Situation after 1~Gyr, when ram pressure reaches its 
maximum, is depicted. Some parcels are allowed to multiply their column 
densities once during the calculation (by a factor of 2.5) in order to 
mimic condensation in situ in the tail (green triangles). Two groups of 
particles develop in the tail -- those retaining their original density 
(magenta dots), and the 'condensed' ones (triangles). The initial 
column density of the parcels is in the calculation limited to 
$100~M_\odot$\,pc$^{-2}$. While some denser parcels could also be 
stripped, they would reach only small distances from the galaxy. 
}\label{Fig005c}
\end{figure}

In order to illustrate the effects of differential ram pressure 
acceleration, we use in Fig.~\ref{Fig005c} simple semi-analytic 
calculations: a set of test particles that represent ISM parcels with a 
range of column densities are initially distributed in a thin (2D) disk 
in a model galaxy potential that experiences an increasing ram pressure 
with a Lorentzian-shaped time profile. Both the model potential and the 
ram pressure pulse are set to approximately correspond to the case of 
D100 in Coma. Equation of motion of the particles under the combined 
effects of the galaxy's potential and (face-on) ram pressure is 
calculated. Fig.~\ref{Fig005c} shows the distribution of velocities and 
distances (relative to the galaxy) that the stripped ISM parcels reach 
after 1~Gyr, at the time when the ram pressure reaches its peak. The 
results are qualitatively consistent with the observations in 
Fig.~\ref{Fig005b} -- a velocity range forms in the distribution of the 
stripped gas parcels as denser clumps are less accelerated. In the 
calculations we moreover mimic simply in-situ condensation of denser 
clumps by multiplying suddenly column density of randomly selected 
parcels. This can reproduce the offset in velocity (triangles vs. dots 
in Fig.~\ref{Fig005b}) -- the 'condensed' parcels decouple from the 
main cloud of stripped parcels (they shift down in the plot). With the 
present calculations we only aim to illustrate the effects of 
differential ram pressure acceleration of the stripped gas components 
with different column densities in the tail, not to accurately model 
the tail of D100.

\section{H$_2$-dominated multi-phase gas tail}\label{multiphase}
D100 is after ESO~137-001 \citep{jachym2014} only the second known 
example of a ram pressure stripped galaxy whose tail shows up 
simultaneously in X-ray, H$\alpha$, and CO emission\footnote{See the 
Discussion section for information about the multiphase tail of the 
Virgo cluster galaxy NGC~4388.}. It is therefore of great interest to 
study how different phases co-exist and evolve in this special 
environment. 

The hot ionized gas is expected to fill up the volume of the tail 
($f\sim 1$). Its total mass is estimated to $\sim 10^8~M_\odot$, which 
corresponds to the observed (bolometric) luminosity of $\sim 1\times 
10^{40}$~erg\,s$^{-1}$ of the soft X-ray emitting gas at temperature of 
$\sim 1.0$~keV \citep[assuming a single-component model with 
metallicity $0.3~Z_\odot$;][]{sanders2014}. The electron density of the 
material is $\sim 8\times 10^{-3}$~cm$^{-3}$ assuming the emission 
comes from a cylinder of 2~kpc radius and 36~kpc in length. 

The warm ionized, H$\alpha$ emitting gas is on the other hand expected 
to have a very low volume filling factor \citep[$f\sim 0.05$; 
see][for details]{jachym2013}. Its total mass is $\sim 2\times 10^8\ 
f^{-1/2}~M_\odot\approx 4.5\times 10^7~M_\odot$, assuming the emission 
fills a 60~kpc$\times$1~kpc cylinder \citep{yagi2007}. 

No \ion{H}{1} was detected in D100 with a $3\sigma$ upper limit of 
$\sim 3\times 10^8~M_\odot$ for the $30''$ beam \citep{bravoalfaro2000, 
bravoalfaro2001} which nearly corresponds to the area of the tail. 
Re-reduction of the original VLA data decreased the $3\sigma$ limit to 
$\sim 0.5\times 10^8~M_\odot$ (Hector Bravo-Alfaro, priv. 
communication). New sensitive VLA observations are forthcoming (PI: M. 
Sun). 

Surprisingly thus, the tail of D100 may be dominated by cold molecular 
component, despite the hot ICM surroundings. The total H$_2$ mass that 
we detected is $\sim 1\times 10^9~M_\odot$, thus by a factor of about 
$5-10$ higher than the mass of ionized hot and warm phases together. 
Given the uncertainty of the $X$-factor (see the discussion in 
Section~\ref{x-factor}), it is possible that the molecular gas fraction 
is somewhat smaller, however, very likely still exceeding the fraction 
of the other gas components. For comparison, in the tail of the Norma 
galaxy ESO~137-001, the amounts of cold molecular and hot ionized gas 
are similar \citep[$\sim 1\times 10^9~M_\odot$;][]{jachym2014, 
sun2006}. 

In numerical simulations of ram pressure stripped gas tails 
\citep{tonnesen2011}, the ratio of warm ($10^4$~K$< T< 10^5$~K) + hot 
($7\times 10^5$~K$< T< 4\times 10^7$~K) to cold ($300$~K$< T< 10^4$~K) 
gas components varies in the range $\sim 2- 9$ for different simulation 
runs (after $\sim 80- 110$~Myr of stripping). In the tail of D100, an 
analogous ratio of warm+hot ionized to cold (molecular+\ion{H}{1}) 
components is thus likely much lower, $\sim 0.15$ (for the standard 
$X$-factor value). Though, more suited simulations would be needed to 
directly compare with the D100 observations, as the CO emitting gas has 
temperatures lower than is the range covered in \citet{tonnesen2011}, 
and the X-ray emitting gas in the D100 tail has temperature of $\sim 
1.2\times 10^7$~K \citep{sanders2014}.

\subsection{Mass gradients along the tail}
The large-scale distribution of the stripped gas along the tail is 
determined by the mass-loss per orbital length, which is given by a 
combination of ram pressure that determines how much gas the galaxy 
loses in a given time, and the orbital velocity that determines over 
which volume the lost gas is spread \citep{roediger2008}. In numerical 
simulations, the mass per orbital length decreases behind the galaxy, 
but at later times saturates (becomes constant) typically at about 
70~kpc from the galaxy \citep[e.g.,][]{roediger2008}. Local physical 
processes, mixing with the surrounding ICM, as well as separation of 
phases drive the evolution of the stripped gas and shape the 
composition of the gas in the tail. 

\begin{table}[t]
\centering
\caption[]
{H$\alpha$ and X-ray fluxes and corresponding masses in IRAM 30m beams.}
\label{TabHa}
\begin{tabular}{lccccc}
\hline
\hline
Source & beam & $F_{\rm H\alpha}$ & $M_{\rm H\alpha}$ & $F_{\rm X-ray}$ & $M_{\rm X-ray}$\\
\noalign{\smallskip}
\hline
\noalign{\smallskip}
D100 & CO10 & $130.5\pm 3.7$ & $-$  & $47.4\pm 15.0$ & $-$ \\
     & CO21 & $129.1\pm 1.1$ & $-$  & $46.6\pm  9.1$ & $-$ \\
T1   & CO10 & $101.9\pm 3.7$ & 14.86& $49.1\pm 14.1$ & 2.3 \\
     & CO21 &   $7.6\pm 0.3$ & 4.05 & $15.8\pm  8.3$ & 0.9 \\
T2   & CO10 &  $11.9\pm 0.3$ & 5.09 & $42.4\pm 15.8$ & 2.1 \\
     & CO21 &   $6.2\pm 0.3$ & 3.66 & $29.9\pm 10.0$ & 1.3 \\
TC2  & CO10 &   $9.4\pm 0.3$ & 4.52 & $69.0\pm 11.6$ & 2.7 \\
     & CO21 &   $4.6\pm 0.2$ & 3.17 & $33.3\pm  8.3$ & 1.3 \\
T3   & CO10 &   $7.9\pm 0.3$ & 4.14 & $62.4\pm 21.6$ & 2.6 \\
     & CO21 &   $4.8\pm 0.2$ & 3.22 & $12.5\pm  8.3$ & 0.8 \\
TC3  & CO10 &   $5.6\pm 0.2$ & 3.47 & $-$            & $-$ \\
     & CO21 &   $2.0\pm 0.1$ & 2.07 & $-$            & $-$ \\
T4   & CO10 &   $5.7\pm 0.2$ & 3.51 &  $3.3\pm  1.7$ & 0.6 \\
     & CO21 &   $4.5\pm 0.2$ & 3.13 &  $6.7\pm  3.3$ & 0.6 \\
\noalign{\smallskip}
\hline
\noalign{\smallskip}
\end{tabular}
\tablecomments{
H$\alpha$ and X-rays fluxes are in $10^{-16}$~erg\,s$^{-1}$\,cm$^{-2}$ 
units, masses in $10^7\,f^{1/2}~M_\odot$ units, where $f$ is the 
respective filling factor. CO(2-1) beam area is $\sim 93.3$~arcsec$^2$; 
CO(1-0) beam area is $\sim 369.9$~arcsec$^2$ ($1/\ln{2}$ factor is not 
included). 
}
\end{table}

\begin{figure}[t]
\centering
\includegraphics[height=0.45\textwidth,angle=270]{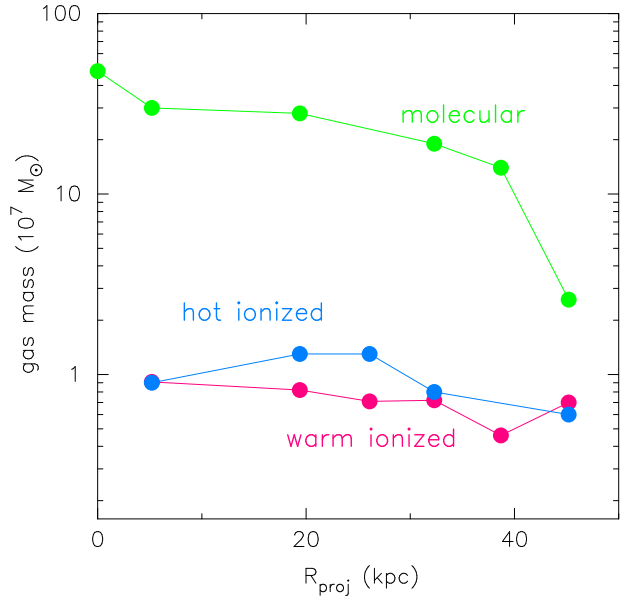}
\caption{
Masses of individual gas components in the observed regions (IRAM 30m 
CO(2-1) beams) as a function of projected distance from the galaxy. A 
volume filling factor of $\sim 0.05$ \citep[see e.g.,][]{jachym2013} 
was applied to the ionized gas (H$\alpha$) masses. We note that the 
lines connecting individual points are intended to lead the eye and do 
not correspond to the actual amount of molecular gas in between the 
observed regions. 
}\label{Fig004}
\end{figure}

Table~\ref{TabHa} gives X-ray and H$\alpha$ fluxes and corresponding 
masses encompassed by our IRAM 30m apertures. In Fig.~\ref{Fig004}, hot 
ionized, warm ionized and molecular gas masses within CO(2-1) apertures 
T1 to T4 are plotted as a function of radial (projected) distance from 
the galaxy. The plot clearly suggests that the balance of the different 
gas phases does not change strongly (within a factor of 2) along most 
of the tail, except its outer parts, where the molecular gas mass drops 
by a factor of $\sim 10$.

To characterize the evolution of the stripped gas we can measure mass 
ratios of individual gas phases. The ratio of molecular to warm + hot 
ionized components decreases from about 17 (in T1) and 13 (in T2, T3) 
to $\sim 2$ (in T4), assuming a constant $X$-factor along the tail. The 
ratio of hot-to-warm ionized (X-ray-to-H$\alpha$) masses is quite 
similar in all tail regions and close to $\sim 1-2$. For comparison, in 
the tail of ESO~137-001, the hot-to-warm gas ratio is also nearly 
constant, $\sim 2$, and the ratio of cold-to-(warm+hot) masses also 
decreases along the tail but in a much less extreme -- from $\sim 3$ in 
the innermost region to $\sim 1$ at 40~kpc distance downstream 
\citep{jachym2014}. The difference could correspond to different 
evolutionary states of stripping of D100 as compared to ESO~137-001.

A possible origin of the observed elevated molecular-to-ionized gas 
ratio could be heating from shocks induced by ram pressure interaction. 
This can create an abundant warm, CO-emitting molecular component in 
the tail. The contribution from shocks is also suggested from optical 
observations that revealed high values of the [\ion{N}{2}]/H$\alpha$ 
ratio of $\sim 0.5- 0.9$ \citep{yagi2007}. Also in the tail of 
ESO~137-001 were measured elevated values of [\ion{N}{2}]/H$\alpha \sim 
0.4$ and [\ion{O}{1}]/H$\alpha \gtrsim 0.2$ \citep[VLT/MUSE 
observations;][]{fossati2016}, as well as in other Coma ram pressure 
stripped galaxies, IC~4040 and RB~199 \citep{yoshida2012}.

\subsection{Correlation of molecular and warm ionized gas phases}
Studying the correlation of individual gas phases in ram pressure 
stripped tails can reveal details about the evolution of the stripped 
ISM and its mixing with the surrounding ICM. A detailed comparison of 
the distributions of different gas phases will be the subject of a 
future study. With the limited level of spatial resolution of our 
current CO observations we focus on the relationship between molecular 
and warm ionized gas in the tail of D100 and in tails of other known 
ram pressure stripped galaxies.

\begin{figure}[t]
\centering
\includegraphics[height=0.45\textwidth,angle=270]{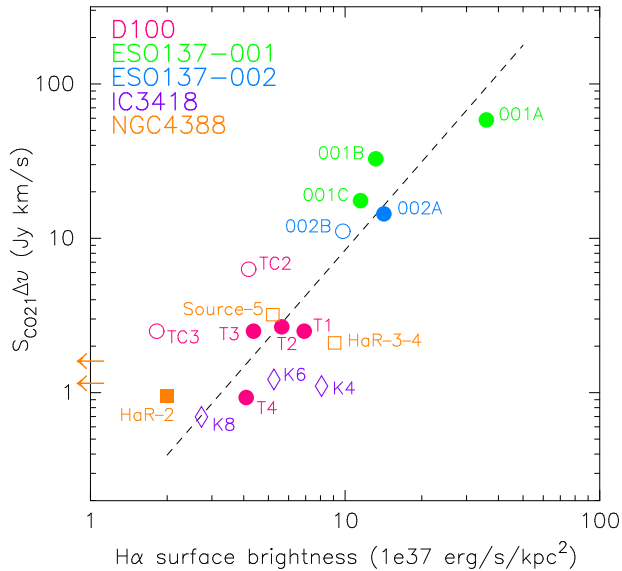}
\caption{
Correlation between CO and H$\alpha$ emission in ram pressure stripped 
gas tails: CO(2-1) integrated flux vs. H$\alpha$ surface brightness 
measured in IRAM 30m apertures ($\rm{HPBW}\sim 5.2$~kpc) over the D100 
tail and APEX apertures ($\rm{HPBW}\sim 8.4$~kpc) over the ESO~137-001 
and ESO~137-002 tails \citep[][J\'achym et al. in prep]{jachym2014}. 
We assumed the widths of the tails of $\sim 2$~kpc (D100), $\sim 
4.8$~kpc (ESO~137-001; or 5.5~kpc at the 001-C region location), and 
$\sim 3$~kpc (ESO~137-002). CO upper limits are marked with open 
symbols. CO(1-0) flux is shown for the TC3 region. Data points from the 
Virgo cluster galaxies IC3418 \citep[diamonds;][]{jachym2013} and 
NGC4388 \citep[squares;][]{verdugo2015} are also shown. The arrows on 
the left edge indicate CO detected regions in NGC~4388 with only weak 
associated H$\alpha$ emission. 
}\label{Fig00ha}
\end{figure}

In Figure~\ref{Fig00ha} the integrated CO fluxes measured in the tails 
of D100 (present work), ESO~137-001 \citep{jachym2014} and ESO~137-002 
(J\'achym et al. in prep.) are plotted as a function of the H$\alpha$ 
surface brightness in the respective tail areas encompassed by IRAM 30m 
or APEX apertures. The tails of the two Norma galaxies are by a factor 
of $\sim 3-10$ brighter than that of D100, both in H$\alpha$ and in CO 
emission. The three data points from ESO~137-001 correspond to APEX 
integrations at $\sim 8$~kpc, 16~kpc, and 40~kpc projected distances 
downstream in the tail, while those from ESO~137-002 are from regions 
in the outer disk and the inner tail (upper limit). There is a 
difference in spatial resolution as the HPBW of the IRAM 30m beam at 
Coma distance is $\sim 5.2$~kpc, while the HPBW of the APEX beam at 
Norma distance is $\sim 8.4$~kpc.

The plot in Fig.~\ref{Fig00ha} reveals a rather tight correlation 
between cold molecular and warm ionized phase over about 2~dex, both in 
CO fluxes and H$\alpha$ surface brightnesses. Molecular gas is the 
primary fuel of star formation. However, the amount of star formation 
in CO-abundant tails is different: there is over 30 \ion{H}{2} regions 
in the tail of ESO~137-001 \citep{sun2007}, possibly only a few 
\ion{H}{2} regions in the tail of D100 (this work), but none were 
revealed in the tail of ESO~137-002 \citep{zhang2013}. The observed 
correlation thus indicates that different excitation processes are 
active in the stripped gas. It is possible that stronger ram pressure 
(either in a richer cluster or in a galaxy that gets closer to the 
cluster center, or in a smaller galaxy) strips more dense gas that can 
form more molecular gas, as well as it produces brighter H$\alpha$ 
emission, not only by star-formation (denser, more compact component) 
but also due to stronger shock ionization (more diffuse component). 
Consequently, stronger ram pressure stripping produces data points at 
the top-right end of the correlation.

In Fig.~\ref{Fig00ha} we further include points corresponding to the CO 
upper limits measured in three H$\alpha$ bright regions in the outer 
tail of the ram pressure stripped Virgo galaxy IC3418 
\citep{jachym2013, fumagalli2011}, and one CO detection and two upper 
limits in the tail of another ram pressure stripped Virgo galaxy 
NGC~4388 \citep{verdugo2015}. While these data points are basically 
well consistent with the revealed correlation, the spatial scale is 
very different -- at the Virgo distance the IRAM 30m CO(2-1) beam HPBW 
is $\sim 0.9$~kpc (or $\sim 1.7$~kpc at CO(1-0)), thus sampling more 
local conditions in the tails. Also, IC3418 is a dwarf galaxy with 
sub-solar metallicity which might be the primary reason for weak CO 
emission, that moreover is in a late stripping state when most of 
molecular gas in the tail would have been consumed by star formation 
\citep{jachym2013, kenney2014}. In NGC~4388 on the other hand, CO 
emission was detected in several more regions not associated with 
strong H$\alpha$ emission ($L_{\rm H\alpha}< 8\times 
10^{35}$~erg\,s$^{-1}$; arrows at the left edge of the plot in 
Fig.~\ref{Fig00ha}).

\subsection{Uncertainty of X-factor}\label{x-factor}
The value of the CO-to-H$_2$ conversion factor is in the D100 tail 
uncertain. Local conditions in the ISM including metallicity, radiation 
field, temperature, density, pressure or dust properties determine the 
value of the conversion between measured CO integrated intensities and 
H$_2$ column densities. 

The environment of the gas-stripped tail is likely distinct from 
typical galactic disks in a number of parameters. For example, the 
volume density of the stripped gas may be lower due to ram pressure 
preferentially stripping lower density gas from the galaxy that 
moreover distributes in an extended three-dimensional tail. Also, the 
relative lack of young stars and thus of strong UV emission suggests 
different chemistry in the tail, possibly driven by shocks and 
turbulence induced by the ram pressure interaction 
\citep[e.g.,][and references therein]{godard2009}. 

In the inner tail of ESO~137-001 elevated ratios of warm-to-cold 
molecular gas of $\gtrsim 0.1$ were observed \citep{sivanandam2010, 
jachym2014}, indicating that some part of CO emission is coming from a 
warmer, more diffuse molecular component. Also in several Virgo 
spirals, higher ratios of warm H$_2$/PAH were found \citep{wong2014}. 
Recent works that analyzed the CO-to-H$_2$ relation in the diffuse 
component of the ISM in the Milky Way \citep{liszt2010, liszt2012, 
romanduval2016} and in nearby galaxies \citep{sandstrom2013} 
surprisingly found that the value of $X$-factor is in the diffuse gas 
similar to that in dense gas. The reason may be that on small spatial 
scales (arcmins in the Galaxy) there are likely large differences in 
the value of the $X$-factor in the diffuse gas between bright CO (low 
$X_{\rm CO}$) and faint CO (high $X_{\rm CO}$) regions, that however on 
large scales average out \citep[see also][for comprehensive 
review]{bolatto2013}. 

While some warm, subthermally excited diffuse component producing 
enhanced CO emission could exist in the D100 tail, there likely are 
also dense, compact regions that re-formed in situ, copying the 
distribution of compact H$\alpha$ regions (see Figure~\ref{Fig003}). 
For example, in the farthest region T4, the measured CO~(2-1)/(1-0) 
line ratio is consistent with a point source distribution. 

We can also estimate the (stellar) metallicity of D100 from the stellar 
mass-metallicity relation of \citet{tremonti2004}. For $M_*= 2.1\times 
10^9~M_\odot$, the predicted median metallicity of D100 is rather high, 
$12+ \log({\rm O/H})\sim 8.7$, i.e., about solar \citep[we assume a 
solar value of 8.69;][]{asplund2009}. While there is no data available 
on the gas metallicity, the above estimate suggests the $X_{\rm CO}$ 
factor is not much affected by metallicity in D100. We expect the 
metallicity of the stripped gas is consistent with the value in the 
main body \citep{fossati2016}. Since D100 is likely in a late 
stripping phase (see Sec.~\ref{SecDiss}), gas originating from outer 
disk regions with potentially lower metallicity due to a radial 
metallicity gradient currently does not contribute to the tail.

In summary, while there probably is an uncertainty in the CO/H$_2$ 
relation of a factor of a few, there is no strong evidence that the 
$X_{\rm CO}$ factor is in the D100 tail systematically lower that 
the standard value. The detected bright CO emission then suggests that 
molecular gas mass dominates the tail by a factor of $\lesssim 5-10$.

\section{Discussion}\label{SecDiss}
Our detection of the abundant molecular gas in the tail of D100 has 
revealed a substantial component of the stripped gas that until now was 
hidden. However, the detection has also raised new questions about the 
evolution of the stripped gas, about the origin of molecular gas in the 
special environment of ram pressure stripped gas tails, as well as 
about how common are the phenomena of molecular gas-rich ram pressure 
stripped tails.

\subsection{Ram pressure stripping origin of the tail}
The 60~kpc multi-wavelength gas tail of D100 has all marks of a ram 
pressure stripping origin as it is gaseous, one-sided, straight and 
well centered on the disk of the galaxy. Moreover the galaxy is 
occurring close to the Coma cluster center, where it is likely 
currently experiencing a peak ram pressure. 

\begin{figure}[t]
\centering
\includegraphics[width=0.445\textwidth]{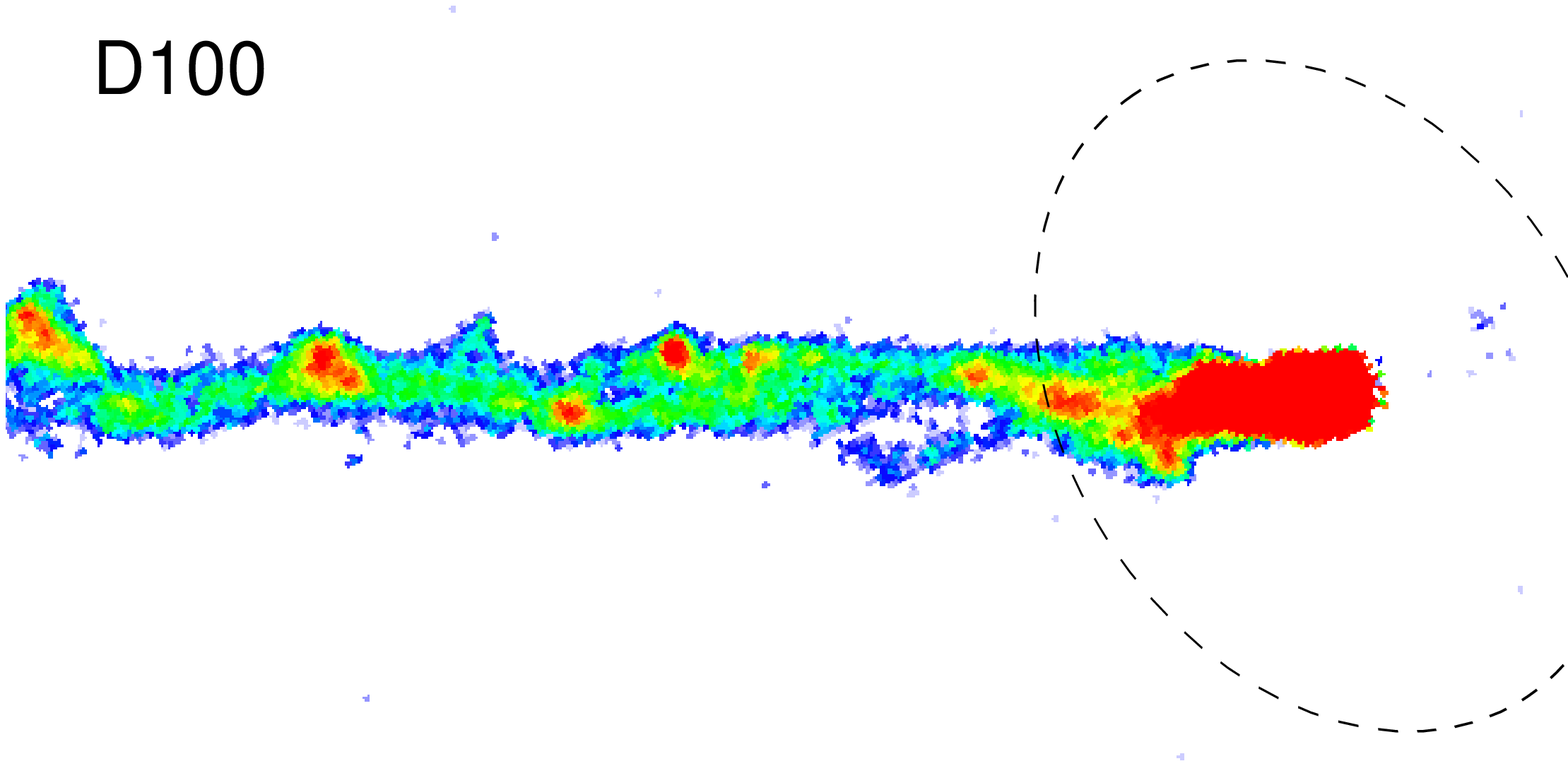}
\includegraphics[width=0.45\textwidth]{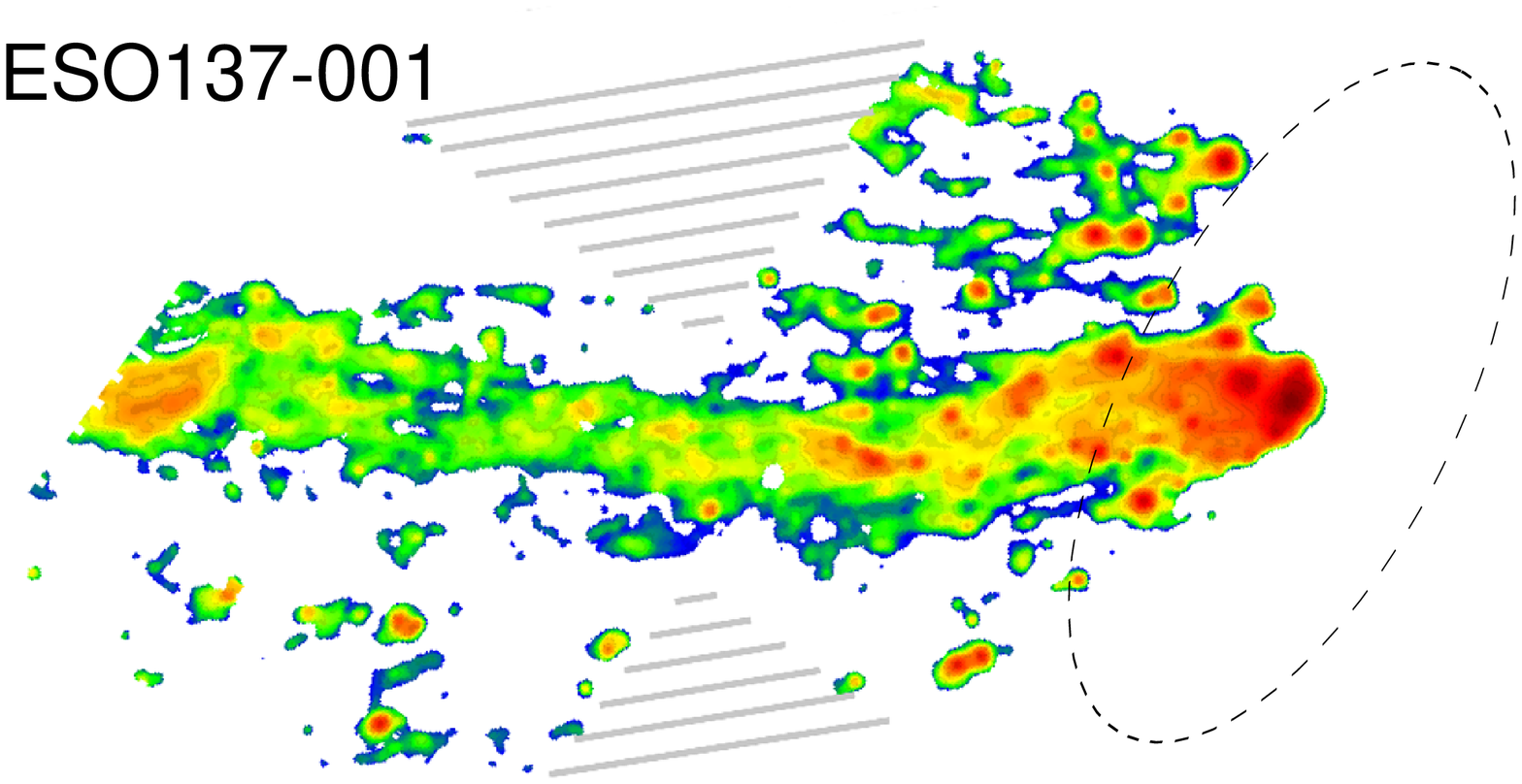}
\caption{
Comparison of inner parts of the tails of D100 \citep[top; 
Subaru;][]{yagi2007} and ESO~137-001 \citep[bottom; 
MUSE;][]{fumagalli2014} as seen in H$\alpha$ emission. The images are 
scaled so that the optical diameters of both galaxies are roughly the 
same ($24''$ in D100 vs. $75''$ in ESO~137-001; marked with dashed 
ellipses). The striking difference in morphology between the two tails 
is clearly visible, mainly due to the presence of a broad component in 
the tail of ESO~137-001. The displayed lengths of the tails correspond 
to $\sim 20$~kpc (D100) and $\sim 40$~kpc (ESO~137-001). The dashed 
areas in the ESO~137-001 image, as well as the continuation of the tail 
to the left-hand-side were not covered by the MUSE observations. 
}\label{FigD100vsESO}
\end{figure}

What makes it clearly stand out of other known cases is its remarkable 
narrowness along its length. In ESO~137-001 for comparison, a wider 
tail of H$\alpha$ regions orphaned from X-ray emission, together with 
the secondary X-ray tail, accompany the main tail. This is clearly 
visible in Fig.~\ref{FigD100vsESO} where we show the inner parts of the 
H$\alpha$ tails of the two galaxies. The images are scaled so that the 
optical diameters of the galaxies match each other. The comparison 
reveals that the tail of D100 has simpler morphology, with much less 
substructure perpendicular to its length, and it is also straighter. On 
the other hand, the two tails bear some similarities -- they are 
brightest in the innermost parts and there are local enhancements along 
the tail with H$\alpha$ peaks separated by relatively diffuse emission 
regions. 

Ram pressure stripping proceeds in a galaxy from the outside in, which 
was confirmed by many observations of truncated HI, H$\alpha$ and/or 
dust disks \citep{koopmann2004}. Consequently the gas stripped at later 
stages is expected to form a narrower tail. Numerical simulations have 
confirmed that small cross-sections lead to less flaring in the tail 
\citep[see e.g.,][their Fig.~8]{roediger2006}. 

The narrow profile of the D100's tail thus suggests that the galaxy is 
in an advanced stripping stage when its stripping radius is small, and 
all ISM from outer disk radii was lost to the intra-cluster space far 
behind the galaxy. Dense (molecular) gas is not expected to be directly 
stripped from the galaxy (see Section~\ref{currentRPS}). A galaxy can 
also have a smaller cross-section if it is moving (near) edge-on 
through the ICM. As we will show later in Section~\ref{SecOrbit}, this 
may be the case for D100, however, this effect would probably be more 
important in an earlier stripping phase. Radiative cooling processes 
were also identified to make for much narrower tails with significantly 
less flaring by reducing the pressure in the tail \citep{tonnesen2010}. 
We also note that the extreme linearity of the tail may point to high 
viscosity of the surrounding ICM that can suppress hydrodynamic 
instabilities that otherwise cause vortices and turbulence 
\citep{roediger2008vis}. Magnetic fields if aligned with the tail can 
further inhibit instabilities and transport processes with the 
surroundings making the tail smoother than it would be in the absence 
of magnetic fields. However, magnetic fields may lead to suppression of 
viscosity, thus the combined effects of the ICM viscosity and magnetic 
fields in the Coma center are likely important for the structure of the 
D100 tail.

\subsubsection{Continuous stripping of nuclear gas?}
The fact that the narrow tail of D100 is not (much) widening suggests 
that the galaxy is heavily stripped and that the stripping radius has 
not changed much over the lifetime of the observed tail. At the same 
time the CO luminosity of the tail is about twice that of the galaxy, 
so a lot of gas has been removed from the galaxy without changing the 
gas stripping radius much. This may be compatible with stripping of 
dense gas from the (circum)nuclear region of the galaxy. D100 contains 
a bar (see the HST image in Figure~\ref{Fig00hst}, right panel) that 
could have driven gas from the outer disk into the inner kiloparsec via 
gravitational torques. This behavior has been observed in barred 
galaxies as indicated from their larger central molecular gas 
concentrations compared to unbarred galaxies \citep{sakamoto1999}. 

The stripping radius of D100 is only $\sim 600$~pc which is about the 
same as the diameter of the bar. When stripping reaches the nuclear 
region ($r<r_{\rm bar}$), the stripping radius does not 
change quickly, both due to large gas concentrations, and the 
gravitational potential that becomes significantly steeper there. In 
many galaxies, especially those with central starbursts, the gas 
surface density increases sharply in the nuclear region, with a scale 
length much smaller than that of the outer disk \citep{jogee2005}. D100 
shows an ongoing nuclear starburst \citep{yagi2007, caldwell1999}. 

Assuming that the stripped gas ($\sim 10^9~M_\odot$) now forming the 
narrow tail originated from an annular region $R= 500- 1000$~pc (which 
corresponds to an area $\sim 10^6$~pc$^2$), we can get a rough estimate 
on the surface density of the original gas in the disk of $\sim 
500~M_\odot$\,pc$^{-2}$. The gas surface density profiles in the 
circumnuclear regions are typically $\sim 100- 300~M_\odot$\,pc$^{-2}$ 
or less at $R\sim 1$~kpc, and $>1000~M_\odot$\,pc$^{-2}$ inside $R\sim 
500$~pc \citep{jogee2005}. It is probable that an analogous galaxy to 
D100 with the same stellar mass but no CN gas concentration would be 
completely stripped at the current D100's location in the cluster, 
possessing only an old detached gas-stripped tail.

\subsubsection{In-situ origin of molecular gas}
Formation and survival of large quantities of (cold) molecular gas in 
the tail is rather surprising given the presence of the surrounding hot 
ICM. It suggests that processes able to heat and disperse the stripped 
cool gas, such as heat conduction, ionizing soft X-ray radiation, 
cosmic rays or turbulence, are not efficient enough to prevent the gas 
from cooling and condensing. For example, assuming the saturated flux 
equations, \citet{tonnesen2011} estimated the efficiency of heat 
conduction to be low in the tails, of the order of $10- 20\%$, 
otherwise cold clouds would quickly evaporate on a timescale of $\sim 
10$~Myr. The viscosity of the ICM is also important for the evolution 
of the stripped gas. With increasing viscosity the stripped galactic 
gas mixes less readily with the ambient ICM. Viscously stripped 
galaxies thus are expected to have unmixed, cool wakes that are also 
X-ray bright \citep{roediger2015}.

A key precondition for H$_2$ formation is the presence of dust in the 
tail, it means its stripping and survival. The HST image of D100 (see 
Fig.~\ref{Fig00hst}, right panel) clearly shows strong dust extinction 
filaments extending from the disk central regions in the direction of 
the tail. This indicates that dust has been stripped to the tail, 
however its survival to large distances from the galaxy is yet to be 
explored. The D100 tail was also covered by \textit{Herschel} 
observations, but only upper limits on FIR emission in the tail can be 
obtained (will be done elsewhere; Sivanadam et al., in prep.). HST 
images of the Virgo galaxies NGC~4522 and NGC~4402 that are being 
stripped clearly show dust being removed from the disk along with the 
gas \citep{abramson2016}.

Given the large amounts of detected molecular gas along the tail, 
including large distances from the galaxy, together with the small 
measured velocity gradients (measured in Section~\ref{kinematic}), 
in-situ molecularization of the stripped material appears to be a 
probable scenario of the origin of the detected gas.
As we will estimate in Sec.~\ref{currentRPS}, ram pressure operating on 
D100 could have directly stripped gas components as dense as $\approx 
50~M_\odot$\,pc$^{-2}$ down to about 1~kpc disk radius. However, due to 
strong internal density gradients in real GMCs reaching values of 
hundreds of $M_\odot$\,pc$^{-2}$, gradual ablation of dense clouds due 
to ram pressure is expected rather than pushing them as a whole. 
Molecular component of the ISM is also expected to be dissociated during 
stripping by shock introduced by ram pressure interaction. Moreover, 
the lifetime of dense molecular clouds is typically $10^6- 10^7$~Myr 
which likely is by at least a factor of 10 shorter than the presumed 
tail age. 

Numerical hydrodynamical simulations suggested that 
overdensities may form from the less dense stripped gas that has cooled 
and condensed in the tail \citep[e.g.,][]{tonnesen2010, tonnesen2012}. 
The timescale for condensation and H$_2$ formation is determined by the 
gas density following an inverse relation \citep{guillard2009}. The 
densest gas lumps in the tail were likely formed from the densest gas 
lumps that were stripped. 
Rather dense clumps stripped from the circumnuclear gas concentration 
of D100, that further were compressed by the surrounding ICM, started to 
cool down efficiently (the cooling time in the tail is 
expected to be rather short). Lack of ambient UV photo-dissociating 
radiation field in the tail further can favor H$_2$ formation on dust 
grains.

\subsection{Star formation in extreme environments}
It is of great interest to study star formation in ram pressure 
stripped gas tails since these are distinct environments from typical 
star-forming regions in disks of spiral galaxies.

\subsubsection{Weak star formation in the tail}
The detailed Subaru Telescope H$\alpha$ image of the tail of D100 shows 
a great deal of substructure in the ionized gas distribution (despite 
its narrowness) with several compact regions visible at different 
distances from the galaxy \citep[][see the image in 
Fig.~\ref{Fig003}]{yagi2010}. The three most compact H$\alpha$ sources 
have also the lowest optical [\ion{N}{2}]/H$\alpha$ ratios in the D100 
tail \citep[in the range $0.44- 0.54$;][their Table~2]{yagi2007}. 
However, these values are too large for typical \ion{H}{2} regions 
\citep[generally $< 0.3$; e.g.,][their Fig. 7]{fossati2016}. Thus, they 
do not look like typical \ion{H}{2} regions formed around star-forming 
clouds. 

We can obtain a rough estimate on the star formation rate (SFR) in 
the tail if we assumed the most compact sources were HII regions. From 
H$\alpha$ fluxes in the three most compact regions identified from the 
H$\alpha$ Subaru image in \citet{yagi2007}\footnote{The sources $\#8$, 
$\#17$, and $\#19$, located at (13:00:10.53, +27:52:10.88) at the edge 
of the CO(1-0) T1 region, (13:00:13.64, +27:52:33.44) in the T3 region, 
and (13:00:15.4, +27:52:45.6) - in the T4 region, respectively.}. The 
corresponding fluxes are $\sim (3.4, 13.0, {\rm and}\ 8.8)\times 
10^{-17}$erg\,s$^{-1}$\,cm$^{-2}$. Using the SFR--$L_{\rm H\alpha}$ 
formula of \citet{kennicutt2012}
\begin{equation}\label{kennicuttHa}
\log{\rm SFR}\ (M_\odot\,{\rm yr}^{-1})= \log{L_{\rm H\alpha}}\ ({\rm
erg\,s^{-1}}) - 41.27,
\end{equation}
the corresponding summed SFR is $\sim 3.9\times 
10^{-3}~M_\odot$\,yr$^{-1}$ assuming a typical 1~mag extinction 
correction for H$\alpha$. For comparison, in the observed regions in 
the tail of ESO~137-001, the summed SFR is by a factor of 10 stronger, 
$\sim 4\times 10^{-2}~M_\odot$\,yr$^{-1}$ \citep{jachym2014}.

A weak UV emission is also seen in the inner $\sim 15$~kpc of the tail 
with {\it GALEX} \citep{smith2010}. The presence of both H$\alpha$ and 
UV emission could suggest star formation has been ongoing in the tail 
within $\sim 10$~Myr (which is the time-scales of the H$\alpha$ 
emission as a tracer of star formation). However, much UV emission 
seems to come from background galaxies that are around the tail and the 
small residual is hard to be isolated for {\it GALEX}'s large PSF. The 
current {\it GALEX} data thus cannot be used to constrain the SFR 
robustly. \citet{smith2010} have also shown a MegaCam archival $u$-band 
deep image with higher resolution than in the {\it GALEX} image. The 
$u$-band continuum shows the tail clearly up to $\sim 30$~kpc. While 
the $u$-band vs. H$\alpha$ overlay does show a lot of correspondence 
between these two tracers, most of the $u$-band emission may not be 
from stars. 

We thus conclude that some star formation may be present in D100's 
tail, though it is weak. Given the wealth of discovered molecular gas, 
the efficiency of star formation must be very low (characterized by 
molecular gas depletion times $\tau_{\rm dep, H_2}= M_{\rm H_2}/{\rm 
SFR}$, that are larger than the Hubble time). Previous observations of 
ESO~137-001 and NGC~4388 indeed suggested that most of molecular gas in 
the stripped gas tails does not form stars and ultimately joins the ICM 
\citep{jachym2014, verdugo2015}.

\subsubsection{Starburst in the disk}
Our proposed scenario of ram pressure stripping of nuclear gas 
concentration is consistent with the starburst observed in the center 
of D100: the nucleus of the galaxy (within $\sim 2''$) exhibits 
starburst characteristics, such as strong emission and strong 
underlying Balmer absorption. Larger radii in the disk ($\sim 3''$) 
show a poststarburst feature (only Balmer absorption) with a quenching 
time between $0- 250$~Myr, depending on the burst strength 
\citep{caldwell1999, yagi2007}. 

We can check what is the depletion timescale of the presumed nuclear 
region in D100. In the main body pointing we detected $\sim 5\times 
10^8~M_\odot$ of molecular gas. The estimated star formation rate in 
the nucleus is $\sim 2.3~M_\odot$\,yr$^{-1}$ (derived from WISE 
band~4), thus $\tau_{\rm dep}\approx 0.22$~Gyr, which is short and 
indeed consistent with a burst.

\subsection{Galaxy's orbit in Coma}\label{SecOrbit}
To better understand the timescale of stripping of D100 we analyze the 
available parameters of its orbit. The galaxy is projected close 
($8.5'\approx 240$~kpc) to the cluster center and its tail points 
nearly perpendicularly to the direction to the cluster center. This 
suggests that D100 is likely near the pericenter of its orbit in 
Coma\footnote{The location and orientation of D100 in Coma is similar 
to that of the completely stripped dwarf IC3418 in Virgo 
\citep{jachym2013, kenney2014}.}. However, the details of the geometry 
of the ram pressure interaction are not immediately clear.

\subsubsection{Constraining the wind angle}
\begin{figure}[]
\centering
\includegraphics[width=0.4\textwidth]{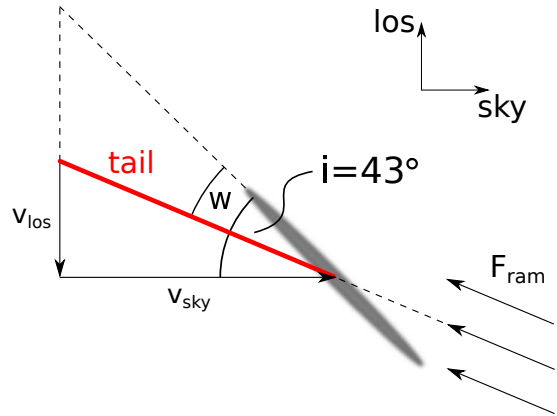}
\caption{
Scheme of the D100 tail geometry: the disk inclination angle is $i\sim 
43^\circ$; the tail extends away from us; the E side of the disk is the 
far side. Consequently, the plane-of-sky orbital velocity component is 
equal to or larger than the l-o-s orbital velocity component. Thus, the 
wind angle ($w$) is close to edge-on (for $\upsilon_{\rm sky}\approx 
\upsilon_{\rm los}$) or $\lesssim 43^\circ$ (for $\upsilon_{\rm sky}\gg 
\upsilon_{\rm los}$), assuming that the tail follows the past orbital 
path of the galaxy. 
}\label{FigScheme}
\end{figure}

We can get an interesting constraint on the ram pressure interaction 
geometry and the 3D velocity of the interaction from the HST image of 
the galaxy (Fig.~\ref{Fig00hst}, right panel). Apparent in the image 
are elongated filaments of dust extinction which can be seen against 
the Eastern disk side. The filaments are elongated in the tail 
direction, and are likely part of the base of the tail. We also know 
that the stripped gas tail must extend in the direction away from us, 
since the radial velocity of D100 is $\sim -1580$~km\,s$^{-1}$ relative 
to the Coma mean \citep[$\sim 6925$~km\,s$^{-1}$;][]{yagi2007}. Since 
the kinematics indicate the tail must extend away from us, and the 
extinction features indicate that the base of the tail is in front of 
the Eastern side of the stellar disk, the Eastern side of the disk must 
be the far side. The geometry is shown in Fig.~\ref{FigScheme} in the 
plane-of-the-sky--line-of-sight view. 

To constrain the wind direction relative to the disk plane (wind 
angle), we first derive the disk inclination angle by inspecting 
isophotes from the HST image. While slight lopsidedness of the disk is 
apparent, consistent values of the major-to-minor axes ratio are 
obtained at several different radii. The corresponding average value of 
the disk inclination is $\sim (43\pm 3)^\circ$. Since the tail is 
extending away from us, and at the same time it is in front of the E 
disk side, the tail angle relative to the disk plane must be in the 
range $\sim 0-43^\circ$. This is illustrated in Fig.~\ref{FigScheme}. 
If we assume that the tail follows the past orbital path, the wind 
angle is equal to the tail angle. Then the above analysis prescribes 
that the plane-of-sky (tangential) orbital velocity component is equal 
to or larger than the radial velocity component of D100. This means 
that the total orbital velocity of D100 is $\gtrsim 
1580\,\sqrt{2}\approx 2200$~km\,s$^{-1}$. 

The system orientation implies that the larger the total orbital 
velocity (and thus the tangential component), the larger the wind 
angle, but it is always smaller than $\sim 43^\circ$. However, for 
reasonable values of the velocity the interaction is close to edge-on: 
for the total orbital velocity of 3000~km\,s$^{-1}$ (or 
4000~km\,s$^{-1}$), the wind angle would be only $\sim 11^\circ$ (or 
$\sim 21^\circ$). 

Numerical simulations have suggested that (close-to) edge-on ram 
pressure stripping is less efficient than face-on or only slightly 
inclined stripping \citep[e.g.,][]{jachym2009, roediger2006}. D100 
however has likely been experiencing very strong (near pericenter) ram 
pressure in which case stripping saturates and the stripping efficiency 
difference between edge-on and face-on stripping disappears 
\citep[cf.,][]{jachym2009}.

\subsubsection{Estimated age of the D100 tail}\label{tailage}
Knowing the disk inclination angle and constraining the tail angle 
relative to the disk allows us to deproject the tail length ($\sim 
60$~kpc in H$\alpha$), as well as the velocity gradient along the tail 
($\sim 150$~km\,s$^{-1}$ in H$\alpha$). For the 3D orbital velocity of 
3000~km\,s$^{-1}$ (or 4000~km\,s$^{-1}$), the deprojected tail length 
would be $\sim 71$~kpc (or $\sim 65$~kpc), and the deprojected velocity 
gradient along the tail $\sim 283$~km\,s$^{-1}$ (or $\sim 
400$~km\,s$^{-1}$). 

We can do a zeroth order estimate on the age of the tail: to cross the 
length of 71~kpc (or 65~kpc) at the velocity of 283~km\,s$^{-1}$ (or 
400~km\,s$^{-1}$), the duration of $\sim 245$~Myr (or $\sim 159$~Myr) 
would be needed. These are probably lower limits as we have neglected 
the fact that gas parcels are from the galaxy ram pressure accelerated 
only gradually. On the other hand, ram pressure changes steeply with 
time as the galaxy orbits through the cluster. 

From the simple estimate of the tail age we can also calculate a zeroth 
order estimate on the average mass loss rate due to stripping. The 
total gas mass of $\sim 1.2\times 10^9~M_\odot$ detected in the tail 
would be stripped from the galaxy over 245~Myr (or 159~Myr) at a high 
rate of $5- 8~M_\odot$\,yr$^{-1}$. 

It is interesting to realize that within the time the tail of the 
observed length has formed, the galaxy has traveled a much larger 
distance through the cluster, possibly exceeding $\sim 0.7- 0.8$~Mpc 
(neglecting the fact that the orbital velocity peaks at the 
pericenter). Thus the environment in which the tail had started to form 
is likely completely disconnected from the galaxy's current location.

\subsubsection{Current ram pressure estimate}\label{currentRPS}
Modeling the ICM distribution in the Coma cluster with a 
$\beta$-profile with parameters given by \citet{mohr1999} and 
\citet{fossati2012}, we can get an upper limit estimate on the ICM 
density at the projected location of D100\footnote{For fully ionized 
gas with primordial abundances (mass of Helium is 0.25 of the total 
mass of the gas), electron and the proton number densities relates to 
the mass density of the ICM as $n_e= 1.167\, n_p$ and $\rho= 1.143\, 
n_e\, {\rm amu}$ \citep[e.g.,][]{pavlovski2008}.} of $\sim 3.3\times 
10^{-27}$~g\,cm$^{-3}$. Assuming the orbital velocity of 
3000~km\,s$^{-1}$ (or 4000~km\,s$^{-1}$), the current ram pressure 
is $\sim 3.0\times 10^{-10}$~dyne\,cm$^{-2}$ (or $\sim 5.3\times 
10^{-10}$~dyne\,cm$^{-2}$). The value is lower by a factor of several 
if the (deprojected) orbit has larger l-o-s distance from the mid-point 
of the cluster. 

For comparison, the current ram pressure acting on D100 may be about 
2-times (or 3.5-times) higher than the estimated current ram pressure 
on ESO~137-001 in the Norma cluster \citep{jachym2014}. Moreover, the 
effects of ram pressure likely have been much stronger on D100 due to 
its $\sim 5$-times lower stellar mass. Solving semi-analytically the 
equation of motion of gas parcels in a D100-like model galaxy 
\citep[see][assuming a $\sim 2\times 10^{11}~M_\odot$ dark matter halo 
and a radial orbit]{jachym2013, jachym2014}, it is probable that the 
gas parcels with column densities up to $\sim 50~M_\odot$\,pc$^{-2}$ 
could be completely (directly) stripped from the galaxy, down to 
$\lesssim 1$~kpc radius.

\subsection{Galaxy transformation watched live}
Several studies of S0 galaxies have suggested that their bulges contain 
younger stellar populations that are also richer in metallicity than 
the surrounding discs \citep[e.g.,][]{johnston2014, cortesi2013}. This 
indicates that the last episode of star formation occurs in the central 
regions of galaxies that are being transformed by cluster environments. 
The starburst observed in the circumnuclear region of D100, that 
otherwise is gas-truncated, is thus consistent with the image of galaxy 
occurring at the late stage of its transformation towards passive S0 
type \citep{johnson2016}. We are lucky enough to witness the ongoing 
transformation, moreover accompanied by the spectacular multi-phase gas 
tail.

\section{Conclusions}
With the IRAM 30m telescope we have discovered large amounts of 
molecular gas in the ram pressure stripped gas tail of the Coma cluster 
galaxy D100 that up to now was known to be bright in H$\alpha$ and 
X-rays. After ESO~137-001 \citep{jachym2014}, this is only the second 
known example of a molecular gas-rich RPS tail. While it is currently 
not clear how common such tails are, it is possible that RPS galaxies 
may be an important source of cold gas (and sometimes stars) to 
intra-cluster space. The main results of our analyses are:

\begin{enumerate}
\item Bright CO emission was detected in several regions along the tail 
of D100 out to nearly 50~kpc distance from the galaxy, with the 
corresponding total H$_2$ mass $\gtrsim10^9~M_\odot$. While the 
value of the $X$-factor in the tail is uncertain and may be locally 
lower by a factor of a few than the standard Galactic value, our 
observations indicate that molecular gas is the dominant mass component 
of the D100 tail. In-situ formation of molecular gas in the stripped 
gas is a preferable scenario of its origin.

\item The extremely narrow morphology of the tail that does not change 
with length indicates that D100 is currently at a late 
evolutionary stage when strong ram pressure has been continuously 
stripping the nuclear dense gas. Along the tail, at a spatial scale of 
$\sim 5$~kpc, the distribution of the stripped gas, including the 
molecular, warm and hot ionized components, is nearly flat, indicating 
that the balance between different gas components does not 
change much. The elevated ratio of molecular-to-ionized gas mass 
measured in the D100 tail may be due to excitation and heating from 
shocks induced by ram pressure interaction. 

\item The kinematics of the stripped gas in the D100 tail reveals an 
offset of $\sim 30$~km\,s$^{-1}$ between denser, more compact gas, 
including CO and compact H$\alpha$ sources, and more diffuse 
H$\alpha$-emitting gas. Such a dynamic separation may be due to 
differential acceleration of the ISM gas phases by ram pressure. 
The stripped gas is along the tail accelerated to only a fraction 
($\sim 10\%$) of the presumed ram pressure wind speed that is 
likely above 3000~km\,s$^{-1}$. The age of 
the (visible) tail is $\sim 200$~Myr. Over that period the galaxy has 
traveled $\sim 800$~kpc through the cluster and thus formation of the 
tail has started in completely disconnected environment. 

\item Some star formation is likely present in the D100 tail but it is 
weak. Given the detected large amounts of molecular gas, the 
efficiency of star formation is very low. This may be due to presumably 
mostly diffuse morphology of the molecular gas. While there are several 
compact regions visible in the H$\alpha$ image of the tail, their 
optical line ratios may be too large for typical HII regions. 

\item The new CO data for D100 in comparison with a limited sample 
of other galaxies with known CO components in ram pressure stripped gas 
tails indicate a rather tight correlation between the CO integrated 
intensity and the H$\alpha$ surface brightness that holds over $\sim 
2$~dex. 
\end{enumerate}

Forthcoming interferometric CO observations of D100 tail will reveal 
the distribution of molecular gas in the tail as well as more details 
about the kinematics of the cold component and allow for a better 
comparison with the kinematics of other gas phases. The present results 
emphasize the importance of multi-wavelength observations for better 
understanding of the fate and evolution of ram pressure stripped gas. 

\section*{Acknowledgments}
This work has been supported by the project LG~14013 of the Ministry of 
Education, Youth and Sports of the Czech republic, the grant project 
15-06012S of the Czech Science Foundation, and the institutional 
research project RVO:67985815. 
M.S. thanks the support provided by the National Aeronautics and Space 
Administration through Chandra Award Number GO6-17111X issued by the 
Chandra X-ray Observatory Center, which is operated by the Smithsonian 
Astrophysical Observatory for and on behalf of the National Aeronautics 
Space Administration under contract NAS8-03060.
We thank Hector Bravo-Alfaro for 
re-reducing existing VLA data. Based in part on data collected at 
Subaru Telescope, which is operated by the National Astronomical 
Observatory of Japan. This research has made use of the NASA/IPAC 
Extragalactic Database (NED) which is operated by the Jet Propulsion 
Laboratory, California Institute of Technology, under contract with the 
National Aeronautics and Space Administration. We further acknowledge 
the usage of the HyperLeda database 
(\url{http://leda.univ-lyon1.fr}).\\

{\it Facilities:} \facility{IRAM 30m}\\

\bibliographystyle{apj}
\bibliography{d100-iram30m}

\end{document}